\documentclass[letterpaper,11pt]{article}

\usepackage{amsmath}
\usepackage{amsfonts}
\usepackage{amssymb}
\usepackage{mathrsfs}
\usepackage{fullpage}
\usepackage{rotating}

\newtheorem{theorem}{Theorem}[section]
\newtheorem{lemma}[theorem]{Lemma}
\newtheorem{definition}[theorem]{Definition}
\newtheorem{problem}[theorem]{Problem}
\newtheorem{corollary}[theorem]{Corollary}
\newtheorem{remark}[theorem]{Remark}

\newcommand{\N}{\mathbb{N}}

\newcommand{\R}{\mathbb{R}}
\newcommand{\BO}{\mathcal{O}}
\newcommand{\A}{\mathcal{A}}
\newcommand{\qed}{\hfill \ensuremath{\Box}}
\newcommand{\proof}[1]{\noindent\textnormal{\textbf{Proof. }}#1\qed}
\newcommand{\Dbi}{\Delta^{\sqcup}}
\newcommand{\Dba}{\Delta^{\circ}}
\newcommand{\sr}[2]{\stackrel{\eqref{#1}}{#2}}

\DeclareMathOperator{\polylog}{polylog}

\title{Tight Bounds for Parallel Randomized Load Balancing}

\author{Christoph Lenzen, Roger Wattenhofer\\
\texttt{\{lenzen,wattenhofer\}@tik.ee.ethz.ch}\\
{\small Computer Engineering and Networks Laboratory (TIK)}\\{\small{ETH
Zurich, 8092 Zurich, Switzerland}}
}

\date{}

\begin{document}

\maketitle

\thispagestyle{empty}

\abstract{We explore the fundamental limits of distributed balls-into-bins
algorithms, i.e., algorithms where balls act in parallel, as separate agents.
This problem was introduced by Adler~et~al., who showed that \emph{non-adaptive}
and \emph{symmetric} algorithms cannot reliably perform better than a maximum bin
load of $\Theta(\log \log n / \log \log \log n)$ within the same number of
rounds. We present an adaptive symmetric algorithm that achieves a bin load of
two in $\log^* n+\BO(1)$ communication rounds using $\BO(n)$ messages in total.
Moreover, larger bin loads can be traded in for smaller time complexities. We
prove a matching lower bound of $(1-o(1))\log^* n$ on the time complexity of
symmetric algorithms that guarantee small bin loads at an asymptotically optimal
message complexity of $\BO(n)$. The essential preconditions of the proof
are $(i)$ a limit of $\BO(n)$ on the total number of messages sent by the
algorithm and $(ii)$ anonymity of bins, i.e., the port numberings of balls are
not globally consistent. In order to show that our technique yields indeed tight
bounds, we provide for each assumption an algorithm violating it, in turn
achieving a constant maximum bin load in constant time.

As an application, we consider the following problem. Given a fully connected
graph of $n$ nodes, where each node needs to send and receive up to $n$ messages,
and in each round each node may send one message over each link, deliver all
messages as quickly as possible to their destinations. We give a simple and
robust algorithm of time complexity $\BO(\log^*n)$ for this task and provide a
generalization to the case where all nodes initially hold arbitrary sets of
messages. Completing the picture, we give a less practical, but asymptotically
optimal algorithm terminating within $\BO(1)$ rounds. All these bounds hold with
high probability.
}

\vspace{\baselineskip}

%
%
%

\newpage

\setcounter{page}{1}

\section{Introduction}

Some argue that in the future understanding parallelism and concurrency will be
as important as understanding sequential algorithms and data structures. Indeed,
clock speeds of microprocessors have flattened about $5$-$6$ years ago. Ever
since, efficiency gains must be achieved by parallelism, in particular using
multi-core architectures and parallel clusters.

Unfortunately, parallelism often incurs a coordination overhead. To be truly
scalable, also coordination must be parallel, i.e., one cannot process
information sequentially, or collect the necessary coordination information at a
single location. A striking and fundamental example of coordination is load
balancing, which occurs on various levels: canonical examples are job assignment
tasks such as sharing work load among multiple processors, servers, or storage
locations, but the problem also plays a vital role in e.g.\ low-congestion
circuit routing, channel bandwidth assignment, or hashing,
cf.~\cite{mitzenmacher01}.

A common archetype of all these tasks is the well-known balls-into-bins problem:
Given $n$ balls and $n$ bins, how can one place the balls into the bins quickly
while keeping the maximum bin load small? As in other areas where centralized
control must be avoided (sometimes because it is impossible), the key to success
is \emph{randomization}. Adler et al.~\cite{adler95} devised parallel randomized
algorithms for the problem whose running times and maximum bin loads are
essentially doubly-logarithmic. They provide a lower bound which is
asymptotically matching the upper bound. However, their lower bound proof
requires two critical restrictions: algorithms must (i) break ties symmetrically
and (ii) be non-adaptive, i.e., each ball restricts itself to a fixed number of
candidate bins before communication starts.

In this work, we present a simple \emph{adaptive} algorithm achieving a maximum
bin load of two within $\log^*n+\BO(1)$ rounds of communication, with high
probability. This is achieved with $\BO(1)$ messages in expectation per ball and
bin, and $\BO(n)$ messages in total. We show that our method is \emph{robust}
to model variations. In particular, it seems that being adaptive helps solving
some practical problems elegantly and efficiently; bluntly, if messages are
lost, they will simply be retransmitted. Moreover, our algorithms can be
generalized to the case where the number of balls differs from the number of
bins.

Complementing this result, we prove that---given the constraints on bin load and
communication complexity---the running time of our algorithm is
$(1+o(1))$-optimal for symmetric algorithms. Our bound necessitates a new proof
technique; it is not a consequence of an impossibility to gather reliable
information in time (e.g.\ due to asynchronicity, faults, or explicitly limited
local views of the system), rather it emerges from bounding the total amount of
communication. Thus, we demonstrate that breaking symmetry to a certain degree,
i.e., reducing entropy far enough to guarantee small bin loads, comes at a cost
exceeding the apparent minimum of $\Omega(n)$ total bits and $\Omega(1)$ rounds.
In this light, a natural question to pose is how much initial entropy is required
for the lower bound to hold. We show that the crux of the matter is that bins are
initially anonymous, i.e., balls do not know globally unique addresses of the
bins. For the problem where bins are consistently labeled $1,\ldots,n$, we give
an algorithm running in constant time that sends $\BO(n)$ messages, yet achieves
a maximum bin load of three. Furthermore, if a small-factor overhead in terms of
messages is tolerated, the same is also possible without a global address space.
Therefore, our work provides a complete classification of the parallel complexity
of the balls-into-bins problem.

Our improvements on parallel balls-into-bins are developed in the context of a
parallel load balancing application involving an even larger amount of
concurrency. We consider a system with $n$ well-connected processors, i.e., each
processor can communicate directly with every other processor.\footnote{This way,
we can study the task of load balancing independently of routing issues.}
However, there is a bandwidth limitation of one message per unit of time on each
connection. Assume that each processor needs to send (and receive) up to $n$
messages, to arbitrary destinations. In other words, there are up to $n^2$
messages that must be delivered, and there is a communication system with a
capacity of $n^2$ messages per time unit. What looks trivial from an
``information theoretic'' point of view becomes complicated if message load is
not well balanced, i.e., if only few processors hold all the $n$ messages for a
single recipient. If the processors knew of each others' intentions, they could
coordinatedly send exactly one of these messages to each processor, which would
subsequently relay it to the target node. However, this simple scheme is
infeasible for reasonable message sizes: In order to collect the necessary
information at a single node, it must receive up to $n^2$ numbers over its $n$
communication links.

In an abstract sense, the task can be seen as consisting of $n$ balls-into-bins
problems which have to be solved \emph{concurrently}. We show that this parallel
load balancing problem can be solved in $\BO(\log^*n)$ time, with high
probability, by a generalization of our symmetric balls-into-bins algorithm.
The resulting algorithm inherits the robustness of our balls-into-bins
technique, for instance it can tolerate a constant fraction of failing edges.
Analogously to the balls-into-bins setting, an optimal bound of $\BO(1)$ on the
time complexity can be attained, however, the respective algorithm is rather
impractical and will be faster only for entirely unrealistic values of $n$. We
believe that the parallel load balancing problem will be at the heart of future
distributed systems and networks, with applications from scientific computing to
overlay networks.

\section{Related Work}
Probably one of the earliest applications of randomized load balancing has been
hashing. In this context, Gonnet \cite{gonnet81} proved that when throwing $n$
balls uniformly and independently at random (u.i.r.) into $n$ bins, the fullest
bin has load $(1+o(1))\log n/\log \log n$ in expectation. It is also common
knowledge that the maximum bin load of this simple approach is $\Theta(\log
n/\log \log n)$ with high probability (w.h.p.)\footnote{I.e., with probability at
least $1-1/n^c$ for a freely choosable constant $c>0$.}~\cite{dubhashi96}.

With the growing interest in parallel computing, since the beginning of the
nineties the topic received increasingly more attention. Karp et
al.~\cite{karp96} demonstrated for the first time that two random choices are
superior to one. By combining two (possibly not fully independent) hashing
functions, they simulated a parallel random access machine (PRAM) on a
distributed memory machine (DMM) with a factor $\BO(\log \log n \log^*n)$
overhead; in essence, their result was a solution to balls-into-bins with maximum
bin load of $\BO(\log\log n)$ w.h.p. Azar et al.~\cite{azar99} generalized their
result by showing that if the balls choose sequentially from $d\geq 2$ u.i.r.\
bins greedily the currently least loaded one, the maximum load is $\log \log
n/\log d+\BO(1)$ w.h.p.\footnote{There is no common agreement on the notion of
w.h.p. Frequently it refers to probabilities of at least $1-1/n$ or $1-o(1)$, as
so in the work of Azar et al.; however, their proof also provides their result
w.h.p.\ in the sense we use throughout this paper.} They prove that this bound is
stochastically optimal in the sense that any other strategy to assign the balls
majorizes\footnote{Roughly speaking, this means that any other algorithm is as
least as likely to produce bad load vectors as the greedy algorithm. An
$n$-dimensional load vector is worse than another, if after reordering the
components of both vectors descendingly, any partial sum of the first $i\in
\{1,\ldots,n\}$ entries of the one vector is greater or equal to the
corresponding partial sum of the other.} their approach. The expected number of
bins each ball queries during the execution of the algorithm was later improved
to $1+\varepsilon$ (for any constant $\varepsilon>0$) by Czumaj and
Stemann~\cite{czumaj01}. This is achieved by placing each ball immediately if the
load of an inspected bin is not too large, rather then always querying $d$ bins.

So far the question remained open whether strong upper bounds can be achieved in
a parallel setting. Adler et al.~\cite{adler95} answered this affirmatively by
devising a parallel greedy algorithm obtaining a maximum load of $\BO(d+\log\log
n/\log d)$ within the same number of rounds w.h.p. Thus, choosing $d\in
\Theta(\log \log n/\log \log \log n)$, the best possible maximum bin load of
their algorithm is $\BO(\log \log n/\log \log \log n)$. On the other hand, they
prove that a certain subclass of algorithms cannot perform better with
probability larger than $1-1/\polylog n$. The main characteristics of this
subclass are that algorithms are \emph{non-adaptive}, i.e., balls have to choose
a fixed number of $d$ candidate bins before communication starts, and
\emph{symmetric}, i.e., these bins are chosen u.i.r. Moreover, communication
takes place only between balls and their candidate bins. In this setting, Adler
et al.\ show also that for any constant values of $d$ and the number of rounds
$r$ the maximum bin load is $\Omega((\log n/\log \log n)^{1/r})$ with constant
probability. Recently, Even and Medina extended their bounds to a larger spectrum
of algorithms by removing some artificial assumptions \cite{even10}. A matching
algorithm was proposed by Stemann~\cite{stemann96}, which for $d=2$ and $r\in
\BO(\log \log n)$ achieves a load of $\BO((\log n/\log \log n)^{1/r})$ w.h.p.;
for $r\in \Theta(\log\log n /\log \log \log n)$ this implies a constantly bounded
bin load. Even and Medina also proposed a 2.5-round ``adaptive''
algorithm~\cite{even09}.\footnote{If balls cannot be allocated, they get an
additional random choice. However, one could also give all balls this additional
choice and let some of them ignore it, i.e., this kind of adaptivity cannot
circumvent the lower bound.} Their synchronous algorithm uses a constant number
of choices and exhibits a maximum bin load of $\Theta(\sqrt{\log n/ \log \log
n})$ w.h.p., i.e., exactly the same characteristics as parallel greedy with 2.5
rounds and two choices. In comparison, within this number of rounds our technique
is capable of achieving bin loads of $(1+o(1))\log \log n/\log \log \log n$
w.h.p.\footnote{This follows by setting $a:=(1+\varepsilon)\log \log n/\log \log
\log n$ (for arbitrary small $\varepsilon>0$) in the proof of
Corollary~\ref{coro:algo_constant}; we get that merely $n/(\log
n)^{1+\varepsilon}$ balls remain after one round, which then can be delivered in
1.5 more rounds w.h.p.\ using $\BO(\log n)$ requests per ball.} See
Table~\ref{table:parallel} for a comparison of our results to parallel
algorithms. Our adaptive algorithms outperform all previous solutions for the
whole range of parameters.

\begin{sidewaystable}
\caption{Comparison of parallel algorithms for $m=n$ balls. Committing balls
into bins counts as half a round with regard to time complexity.}
\label{table:parallel}
\begin{center}
\begin{tabular}{|c|c|c|c|c|c|c|}\hline
algorithm  & symmetric & adaptive & choices & rounds & maximum bin load &
messages\\ \hline\hline
naive \cite{gonnet81}  & yes & no  & $1$ & $0.5$ & $\BO\big(\frac{\log n}{\log
\log n}\big)$ & $n$\\[.5ex]
par.\ greedy \cite{adler95}  & yes & no  & $2$ & 2.5 &
$\BO\Big(\sqrt{\frac{\log n}{\log \log n}}\,\Big)$ & $\BO(n)$\\
par.\ greedy \cite{adler95}  & yes & no  &
$\Theta\big(\frac{\log \log n}{\log\log\log n}\big)$
& $\Theta\big(\frac{\log \log n}{\log\log\log n}\big)$
& $\BO\big(\frac{\log \log n}{\log\log\log n}\big)$
& $\BO\big(\frac{n\log \log n}{\log\log\log n}\big)$\\
collision \cite{stemann96}  & yes & no  & $2$ & $r+0.5$ &
$\BO\Big(\big(\frac{\log n}{\log \log n}\big)^{1/r}\Big)$ & $\BO(n)$\\
$\A_b^2$  & yes & yes  & $\BO(1)$ (exp.) & $\log^*n+\BO(1)$ & $2$ & $\BO(n)$\\
$\A_b(r)$  & yes & yes  & $\BO(1)$ (exp.) & $r+\BO(1)$ &
$\frac{\log^{(r)}n}{\log^{(r+1)}n}+r+\BO(1)$ & $\BO(n)$\\
$\A_c(l)$ & yes & yes & $\BO(l)$
(exp.) & $\log^* n-\log^* l+\BO(1)$ & $\BO(1)$ & $\BO(ln)$\\
$\A(\sqrt{\log n})$ & no & yes & $\BO(1)$ (exp.) &
$\BO(1)$ & $3$ & $\BO(n)$\\
\hline
\end{tabular}
\end{center}
\end{sidewaystable}

Given the existing lower bounds, since then the only possibility for further
improvement has been to search for non-adaptive or asymmetric algorithms.
V\"ocking~\cite{voecking03} introduced the sequential ``always-go-left''
algorithm which employs asymmetric tie-breaking in order to improve the impact
of the number of possible choices $d$ from logarithmic to linear. Furthermore,
he proved that dependency of random choices does not offer asymptotically
better bounds. His upper bound holds also true if only two bins are chosen
randomly, but for each choice $d/2$ consecutive bins are
queried~\cite{kenthapadi06}. Table \ref{table:sequential} summarizes sequential
balls-into-bins algorithms. Note that not all parallel algorithms can also be
run sequentially.\footnote{Stemann's collision protocol, for instance, requires
bins to accept balls only if a certain number of pending requests is not
exceeded. Thus the protocol cannot place balls until all random choices are
communicated.} However, this is true for our protocols; our approach translates
to a simple sequential algorithm competing in performance with the best known
results~\cite{czumaj01,voecking03}. This algorithm could be interpreted as a
greedy algorithm with $d=\infty$.

Most of the mentioned work considers also the general case of $m\neq n$. If $m>
n$, this basically changes expected loads to $m/n$, whereas values considerably
smaller than $n$ (e.g.~$n^{1-\varepsilon}$) admit constant maximum bin load. It
is noteworthy that for $d\geq 2$ the imbalance between the most loaded bins and
the average load is $\BO(\log \log n/\log d)$ w.h.p.\ irrespective of $m$.
Recently, Peres et al.~\cite{peres10} proved a similar result for the case where
``$d=1+\beta$'' bins are queried, i.e., balls choose with constant probability
$\beta\in (0,1)$ the least loaded of two bins, otherwise uniformly at random. In
this setting, the imbalance becomes $\Theta((\log n)/\beta)$ w.h.p.

In addition, quite a few variations of the basic problem have been studied.
Since resources often need to be assigned to dynamically arriving tasks,
infinite processes have been considered
(e.g.~\cite{azar99,czumaj01,mitzenmacher96,mitzenmacher98b,mitzenmacher98a,stemann96,voecking03}).
In~\cite{mitzenmacher02a} it is shown that, in the sequential setting,
memorizing good choices from previous balls has similar impact as increasing
the number of fresh random choices. Awerbuch et al.~\cite{awerbuch95} studied
arbitrary $L_p$ norms instead of the maximum bin load (i.e., the $L_{\infty}$
norm) as quality measure, showing that the greedy strategy is $p$-competitive
to an offline algorithm. Several works addressed weighted balls
(e.g.~\cite{berenbrink08,berenbrink97,koutsoupias03,peres10,talwar07}) in order
to model tasks of varying resource consumption. The case of heterogeneous bins
was examined as well~\cite{wieder07}. In recent years, balls-into-bins has
also been considered from a game theoretic point of
view~\cite{berenbrink07,kleinberg09}.

\begin{table*}
\caption{Comparison of sequential algorithms for $m=n$
balls.}\label{table:sequential}
\begin{center}
\begin{tabular}{|c|c|c|c|c|c|}\hline
algorithm & sym. & adpt. & choices & max. bin load & bin queries\\
\hline\hline
naive \cite{gonnet81} & yes & no & $1$ & $\BO\big(\frac{\log n}{\log \log
n}\big)$ & $n$\\
greedy \cite{azar99} & yes & no & $d\geq 2$ & $\frac{\log \log n}{\log
d}+\BO(1)$ & $\BO(d n)$\\
always-go-left \cite{voecking03} & no & no & $d\geq 2$
& $\BO\big(\frac{\log \log n}{d}\big)$ & $\BO(d n)$\\
adpt.\ greedy \cite{czumaj01} & yes & yes & $1+o(1)$ (exp.); at most
$d\geq 2$ & $\BO\big(\frac{\log \log n}{\log d}\big)$ & $(1+o(1))n$\\
$\A_{\mbox{seq}}$ & yes & yes & $\BO(1)$ (exp.) & $2$ & $(2+o(1))n$\\
\hline
\end{tabular}
\end{center}
\end{table*}

Results related to ours have been discovered before for hashing problems. A
number of publications presents algorithms with running times of $\BO(\log^* n)$
(or very close) in PRAM models~\cite{bast91,gil96,matias91,meyer96}. At the heart
of these routines as well as our balls-into-bins solutions lies the idea to use
an in each iteration exponentially growing share of the available resources to
deal with the remaining keys or bins, respectively. Implicitely, this approach
already occured in previous work by Raman~\cite{raman90}. For a more detailed
review of these papers, we refer the interested reader to~\cite{hagerup92}.
Despite differences in the models, our algorithms and proofs exhibit quite a few
structural similarities to the ones applicable to hashing in PRAM models. From
our point of view, there are two main differences distinguishing our upper bound
results on symmetric algorithms. Firstly, the parallel balls-into-bins model
permits to use the algorithmic idea in its most basic form. Hence, our
presentation focuses on the properties decisive for the $\log^* n+\BO(1)$
complexity bound of the basic symmetric algorithm. Secondly, our analysis shows
that the core technique is highly robust and can therefore tolerate a large
number of faults.

The lower bound by Adler~et~al.\ (and the generalization by Even and Medina) is
stronger than our lower bound, but it applies to algorithms which are severely
restricted in their abilities only. Essentially, these restrictions uncouple the
algorithm's decisions from the communication pattern; in particular,
communication is restricted to an initially fixed random graph, where each ball
contributes $d$ edges to u.i.r.\ bins. This prerequisite seems reasonable for
systems where the initial communication overhead is large. In general, we find it
difficult to motivate that a non-constant number of communication rounds is
feasible, but an initially fixed set of bins may be contacted only. In contrast,
our lower bound also holds for adaptive algorithms. In fact, it even holds for
algorithms that allow for address forwarding, i.e., balls may contact any bin
deterministically after obtaining its globally unique address.\footnote{This
address is initially known to the respective bin only, but it may be forwarded
during the course of an algorithm.} In other words, it arises from the assumption
that bins are (initially) anonymous (cf.~Problems~\ref{prob:sbib}
and~\ref{prob:acbib}), which fits a wide range of real-world systems.

Like Linial in his seminal work on 3-coloring the ring~\cite{linial92}, we attain
a lower bound of $\Omega(\log^* n)$ on the time required to solve the task
efficiently. This connection is more than superficial, as both bounds essentially
arise from a symmetry breaking problem. However, Linial's argument uses a highly
symmetric ring topology.\footnote{This general approach to argue about a simple
topology has been popular when proving lower
bounds~\cite{czygrinow08,lenzen08,naor91}.} This is entirely different from our
setting, where any two parties may potentially exchange information. Therefore,
we cannot argue on the basis that nodes will learn about a specific subset of the
global state contained within their local horizon only. Instead, the random
decisions of a balls-into-bins algorithm define a graph describing the flow of
information. This graph is not a simple random graph, as the information gained
by this communication feeds back to its evolution over time, i.e., future
communication may take the local topology of its current state into account.

A different lower bound technique is by Kuhn~et~al.~\cite{kuhn10lowertech}, where
a specific locally symmetric, but globally asymmetric graph is constructed to
render a problem hard. Like in our work, \cite{kuhn10lowertech} restricts its
arguments to graphs which are locally trees. The structure of the graphs we
consider imposes to examine subgraphs which are trees as well; subgraphs
containing cycles occur too infrequently to constitute a lower bound. The bound
of $\Omega(\log^* n)$ from~\cite{gil96}, applicable to hashing in a certain
model, which also argues about trees, has even more in common with our result.
However, neither of these bounds needs to deal with the difficulty that the
algorithm may influence the evolution of the communication graph in a complex
manner. In~\cite{kuhn10lowertech}, input and communication graph are identical
and fixed; in~\cite{gil96}, there is also no adaptive communication pattern, as
essentially the algorithm may merely decide on how to further separate elements
that share the same image under the hash functions applied to them so far.

Various other techniques for obtaining distributed lower bounds
exist~\cite{fich03,lynch89}, however, they are not related to our work. If
graph-based, the arguments are often purely information theoretic, in the sense
that some information must be exchanged over some bottleneck link or node in a
carefully constructed network with diameter larger than
two~\cite{lotker06,peleg99}. In our setting, such information theoretic lower
bounds will not work: Any two balls may exchange information along $n$
edge-disjoint paths of length two, as the graph describing which edges could
\emph{potentially} be used to transmit a message is complete bipartite. In some
sense, this is the main contribution of this paper: We show the existence of a
coordination bottleneck in a system without a physical bottleneck.


%

The remainder of this technical report is organized as follows. In
Section~\ref{sec:communication}, we state and solve the aforementioned load
balancing problem in a fully connected system. The discussion of the related
symmetric balls-into-bins algorithm $\A_b$ is postponed to
Section~\ref{sec:balls_bins}, as the more general proofs from
Section~\ref{sec:communication} permit to infer some of the results as
corollaries. After discussing $\A_b$ and its variations, we proceed by developing
the matching lower bound in Section~\ref{sec:lower}. Finally, in
Section~\ref{sec:upper}, we give algorithms demonstrating that if any of the
prerequisites of the lower bound does not hold, constant-time constant-load
solutions are feasible.

\section{Preliminary Statements}\label{sec:basics}
Our analysis requires some standard definitions and tools, which are summarized
in this section.
\begin{definition}[Uniformity and Indepence]
The (discrete) random variable $X:\Omega\to S$ is called \emph{uniform}, if
$P[X=s_1]=P[X=s_2]$ for any two values $s_1,s_2\in S$. The random variables
$X_1:\Omega_1\to S_1$ and $X_2:\Omega_2\to S_2$ are \emph{independent}, if for
any $s_1\in S_1$ and $s_2\in S_2$ we have $P[X_1=s_1]=P[X_1=s_1|X_2=s_2]$ and
$P[X_2=s_2]=P[X_2=s_2|X_1=s_1]$. A set $\{X_1,\ldots,X_N\}$ of variables is
called independent, if for any $i\in \{1,\ldots,N\}$, $X_i$ is independent from
the variable $(X_1,\ldots,X_{i-1},X_{i+1},\ldots,X_n)$, i.e., the variable
listing the outcomes of all $X_j\neq X_i$. The set $\{X_1,\ldots,X_N\}$ is
\emph{uniformly and independently at random (u.i.r.)} if and only if it is
independent and consists of uniform random variables. Two sets of random
variables $X=\{X_1,\ldots,X_N\}$ and $Y=\{Y_1,\ldots,Y_M\}$ are independent if
and only if all $X_i\in X$ are independent from $(Y_1,\ldots,Y_M)$ and all
$Y_j\in Y$ are independent from $(X_1,\ldots,X_N)$.
\end{definition}

We will be particularly interested in algorithms which almost guarantee certain
properties.
\begin{definition}[With high probability (w.h.p.)]
We say that the random variable $X$ attains values from the set $S$ \emph{with
high probability}, if $P[X\in S]\geq 1-1/n^c$ for an arbitrary, but fixed
constant $c>0$. More simply, we say $S$ occurs w.h.p.
\end{definition}
The advantage of this stringent definition is that any polynomial number of
statements that individually hold w.h.p., also hold w.h.p.\ in conjunction.
Throughout this paper, we will use this lemma implicitly, as we are always
interested in sets of events whose sizes are polynomially bounded in $n$.
\begin{lemma}\label{lemma:transitivity}
Assume that statements $S_i$, $i\in \{1,\ldots,N\}$, hold w.h.p., where $N\leq
n^d$ for some constant $d$. Then $S:=\bigwedge_{i=1}^N S_i$ occurs w.h.p.
\end{lemma}
\proof{The $S_i$ hold w.h.p., so for any fixed constant $c>0$ we may choose
$c':=c+d$ and have $P(S_i)\geq 1-1/n^{c'}\geq 1-1/(N n^c)$ for all $i\in
\{1,\ldots,N\}$. By the union bound this implies $P[S]\geq 1-\sum_{i=1}^N
P[\overline{S_i}]\geq 1-1/n^c$.}

Frequently w.h.p.\ results are deduced from Chernoff type bounds, which provide
exponential probability bounds regarding sums of Bernoulli variables. Common
formulations assume independence of these variables, but the following
more general condition is sufficient.
\begin{definition}[Negative Association]
The random variables $X_i$, $i\in\{1,\ldots,N\}$, are \emph{negatively
associated} if and only if for all disjoint subsets $I,J\subseteq
\{1,\ldots,N\}$ and all functions $f:\R^{|I|}\to \R$ and $g:\R^{|J|}\to \R$
that are either increasing in all components or decreasing in all components we
have
\begin{equation*}
E(f(X_i,i\in I)\cdot g(X_j,j\in J))\leq E(f(X_i,i\in I))\cdot E(g(X_j,j\in
J)).
\end{equation*}
\end{definition}
Note that independence trivially implies negative association, but not vice
versa. Using this definition, we can state a Chernoff bound suitable to our
needs.
\begin{theorem}[Chernoff's Bound]\label{theorem:chernoff}
Let $X:=\sum_{i=1}^N X_i$ be the sum of $N$ negatively associated
Bernoulli variables $X_i$. Then, w.h.p.,
\begin{itemize}
  \item [(i)] $E[X]\in \BO(\log n) \Rightarrow X\in \BO(\log n)$
  \item [(ii)] $E[X]\in \BO(1) \Rightarrow X\in \BO\left(\frac{\log
  n}{\log \log n}\right)$
  \item [(iii)] $E[X]\in \omega(\log n) \Rightarrow X\in (1\pm o(1))E[X]$.
\end{itemize}
\end{theorem}

In other words, if the expected value of a sum of negatively associated
Bernoulli variables is small, it is highly unlikely that the result will be of
more than logarithmic size, and if the expected value is large, the outcome will
almost certainly not deviate by more than roughly the square root of the
expectation. In the forthcoming, we will repeatedly make use of these basic
observations.

In order to do so, techniques to prove that sets of random variables are
negatively associated are in demand. We will rely on the following results
of Dubhashi and Ranjan \cite{dubhashi96}.
\begin{lemma}\label{lemma:negative_association}\
\begin{itemize}
  \item[(i)] If $X_1,\ldots,X_{N}$ are Bernoulli variables satisfying
  $\sum_{i=1}^{N}X_i= 1$, then $X_1,\ldots,X_{N}$ are negatively
  associated.
  \item[(ii)] Assume that $X$ and $Y$ are negatively associated sets of random
  variables, and that $X$ and $Y$ are mutually independent. Then $X\cup Y$ is 
  negatively associated.
  \item[(iii)] Suppose $\{X_1,\ldots,X_{N}\}$ is negatively associated.
  Given $I_1,\ldots,I_k\subseteq\{1,\ldots,N\}$, $k\in \N$, and
  functions $h_j:\R^{|I_j|}\to \R$, $j\in \{1,\ldots,k\}$, that are either all
  increasing or all decreasing, define $Y_j:=h_j(X_i,i\in I_j)$. Then
  $\{Y_1,\ldots,Y_k\}$ is negatively associated.
\end{itemize}
\end{lemma}

This lemma and Theorem \ref{theorem:chernoff} imply strong bounds on the
outcome of the well-known balls-into-bins experiment.
\begin{lemma}\label{lemma:balls_bins_negative}
Consider the random experiment of throwing $M$ balls u.i.r.\ into $N$ bins.
Denote by $Y_i^k$, $i\in \{1,\ldots,N\}$, the set of Bernoulli variables being
$1$ if and only if at least (at most) $k\in \N_0$ balls end up in bin $i\in
\{1,\ldots,N\}$. Then, for any $k$, the set $\{Y_i^k\}_{i\in \{1,\ldots,N\}}$ is
negatively associated.
\end{lemma}
\proof{Using Lemma \ref{lemma:negative_association}, we can pursue the
following line of argument:
\begin{enumerate}
  \item For each ball $j\in \{1,\ldots,M\}$, the Bernoulli variables
  $\{B_j^i\}_{i\in \{1,\ldots,N\}}$ which are $1$ exactly if ball $j$ ends up
  in bin $i$, are negatively associated (Statement~$(i)$ from Lemma
  \ref{lemma:negative_association}).
  \item The whole set $\{B_j^i\,|\,i\in \{1,\ldots,N\}\wedge j\in
  \{1,\ldots,M\}\}$ is negatively associated (Statement~$(ii)$ of
  Lemma~\ref{lemma:negative_association}).
  \item The sets $\{Y_i^k\}_{i\in \{1,\ldots,N\}}$ are for each $k\in \N_0$
  negatively associated (Statement~$(iii)$ of
  Lemma~\ref{lemma:negative_association}).
\end{enumerate}}

The following special case will be helpful in our analysis.
\begin{corollary}\label{coro:empty_bins}\ \\
Throw $M\leq N\ln N /(2\ln\ln n)$ balls u.i.r.\ into $N$ bins. Then w.h.p.\
$(1\pm o(1))N e^{-M/N}$ bins remain empty.
\end{corollary}
\proof{The expected number of empty bins is $N(1-1/N)^M$. For $x\geq 1$ and
$|t|\leq x$ the inequality $(1-t^2/x)e^t\leq (1+t/x)^x\leq e^t$ holds. Hence,
with $t=-M/N$ and $x=M$ we get
\begin{equation*}
(1-o(1))e^{-M/N}\ni \left(1-\frac{(M/N)^2}{N}\right)e^{-M/N}
\leq \left(1-\frac{1}{N}\right)^M\leq e^{-M/N}.
\end{equation*}
Due to the upper bound on $M$, we have $N e^{-M/N}\geq (\ln n)^2\in
\omega(\log n)$. Lemma \ref{lemma:balls_bins_negative} shows that we can apply
Theorem \ref{theorem:chernoff} to the random variable counting the number of empty
bins, yielding the claim.}

Another inequality that yields exponentially falling probability bounds is
typically referred to as Azuma's inequality.
\begin{theorem}[Azuma's Inequality]\label{theorem:azuma}
Let $X$ be a random variable which is a function of independent random
variables $X_1,\ldots,X_{N}$. Assume that changing the value of a single
$X_i$ for some $i\in \{1,\ldots,N\}$ changes the outcome of $X$ by at most
$\delta_i\in \R^+$. Then for any $t\in \R^+_0$ we have
\begin{eqnarray*}
P\big[|X-E[X]|>t\big]&\leq & 2e^{-t^2/\left(2\sum_{i=1}^{N}\delta_i^2\right)}.
\end{eqnarray*}
\end{theorem}

\section{Parallel Load Balancing}\label{sec:communication}
In this section, we examine the problem of achieving as low as possible
congestion in the complete graph $K_n$ if links have uniform capacity. In order
to simplify the presentation, we assume that all loops $\{v,v\}$, where $v\in
V:=\{1,\ldots,n\}$, are in the edge set, i.e., nodes may ``send messages to
themselves''. All nodes have unique identifiers, that is, $v\in V$ denotes both
the node $v$ and its identifier. We assume that communication is synchronous and
reliable.\footnote{This is convenient for ease of presentation. We will see later
that both assumptions can be dropped.} During each synchronous round, nodes may
perform arbitrary local computations, send a (different) message to each other
node, and receive messages.

We will prove that in this setting, a probabilistic algorithm enables nodes to
fully exploit outgoing and incoming bandwidth (whichever is more restrictive) with
marginal overhead w.h.p. More precisely, we strive for enabling nodes to freely
divide the messages they can send in each round between all possible destinations
in the network. Naturally, this is only possible to the extent dictated by the
capability of nodes to receive messages in each round, i.e., ideally the amount
of time required would be proportional to the maximum number of messages any
node must send or receive, divided by $n$.

This leads to the following problem formulation.
\begin{problem}[Information Distribution Task]\label{prob:idt}
Each node $v\in V$ is given a (finite) set of messages
\begin{equation*}
{\cal S}_v=\{m_v^i\,|\,i\in I_v\}
\end{equation*}
with destinations $d(m_v^i)\in V$, $i\in I_v$. Each such message explicitly
contains $d(m_v^i)$, i.e., messages have size $\Omega(\log n)$. Moreover,
messages can be distinguished (e.g., by also including the sender's identifier
and the position in an internal ordering of the messages of that sender). The
goal is to deliver all messages to their destinations, minimizing the total
number of rounds. By
\begin{equation*}
{\cal R}_v:=\left\{m_w^i\in \bigcup_{w\in V}{\cal
S}_w\,\Bigg|\,d(m_w^i)=v\right\}
\end{equation*}
we denote the set of messages a node $v\in V$ shall receive. We abbreviate
$M_s:=\max_{v\in V}|{\cal S}_v|$ and $M_r:=\max_{v\in V}|{\cal R}_v|$, i.e.,
the maximum numbers of messages a single node needs to send or receive,
respectively.
\end{problem}

We will take particular interest in a special case.

\begin{problem}[Symmetric Information Distribution Task]\label{prob:sidt}
An instance of Problem \ref{prob:idt} such that for all $v\in V$ it holds that
$|{\cal S}_v|=|{\cal R}_v|=n$, i.e., all nodes have to send and receive exactly
$n$ messages, is called \emph{symmetric information distribution task}.
\end{problem}

\subsection{Solving the Symmetric Information Distribution Task}
In order to achieve small time bounds, we will rely on temporary replication of
messages. Nodes then deliver (at most) a constant number of copies to each
recipient and restart the procedure with the messages of which no copy arrived at
its destination. However, a large number of duplicates would be necessary to
guarantee immediate success for all messages. This is not possible right from the
start, as the available bandwidth would be significantly exceeded.
Therefore, it seems to be good advice to create as many copies as possible
without causing too much traffic. This inspires the following algorithm.

At each node $v\in V$, algorithm $\A_s$ running on $K_n$ executes the following
loop until it terminates:
\begin{enumerate}
  \item Announce the number of currently held messages to all other nodes. If
  no node has any messages left, terminate.
  \item Redistribute the messages evenly such that all nodes store (up to one)
  the same amount.\footnote{Since all nodes are aware of the number of messages
  the other nodes have, this can be solved deterministically in one round: E.g.,
  order the messages $m_v^i$, $v\in V$, $i\in I_v$, according to $m_v^i<m_w^j$
  if $v<w$ or $v=w$ and $i<j$, and send the $k^{th}$ message to node $k$
  mod $n$; all nodes can compute this scheme locally without communication.
  Since no node holds more than $n$ messages, one round of communication is
  required to actually move the messages between nodes.}
  \item Announce to each node the number of messages for it you currently hold.
  \item Announce the total number of messages destined for you to all other
  nodes.
  \item Denoting by $M_r'$ the maximum number of messages any node still
  needs to receive, create $k:=\left\lfloor n/M_r'\right\rfloor$ copies of each
  message. Distribute these copies uniformly at random among all nodes, but
  under the constraint that no node gets more than one of the
  duplicates.\footnote{Formally: Enumerate the copies arbitrarily and send the
  $i^{th}$ copy to node $\sigma(i)$, where $\sigma\in S_n$ is a permutation of
  $\{1,\ldots,n\}$ drawn uniformly at random.}
  \item To each node, forward one copy of a message destined for it (if any has
  been received in the previous step; if multiple copies have been received,
  any choice is feasible) and confirm the delivery to the previous sender.
  \item Delete all messages for which confirmations have been received and all
  currently held copies of messages.
\end{enumerate}
\begin{remark}
Balancing message load and counting the total number of messages (Steps~1 to~4)
is convenient, but not necessary. We will later see that it is possible to
exploit the strong probabilistic guarantees on the progress of the algorithm in
order to choose proper values of $k$.
\end{remark}
\begin{definition}[Phases]
We will refer to a single execution of the loop, i.e., Steps 1 to 7, as a
\emph{phase}.
\end{definition}

Since in each phase some messages will reach their destination, this algorithm
will eventually terminate. To give strong bounds on its running time, however, we
need some helper statements. The first lemma states that in Steps~5 and 6 of
the algorithm a sufficiently large uniformly random subset of the duplicates will
be received by their target nodes.
\begin{lemma}\label{lemma:sufficient_messages}
Denote by ${\cal C}_v$ the set of copies of messages for a node $v\in V$
that $\A_s$ generates in Step~5 of a phase. Provided that $|{\cal C}_v|\in
\omega(\log n)$ and $n$ is sufficiently large, w.h.p.\ the set of messages $v$
receives in Step~6 contains a uniformly random subset of ${\cal C}_v$ of size
at least $|{\cal C}_v|/4$.
\end{lemma}
\proof{Set $\lambda:=|{\cal C}_v|/n\leq 1$. Consider the random experiment where
$|{\cal C}_v|=\lambda n$ balls are thrown u.i.r.\ into $n$ bins. We make a
distinction of cases. Assume first that $\lambda\in [1/4,1]$. Denote for $k\in
\N_0$ by $B_k$ the random variable counting the number of bins receiving exactly
$k$ balls. According to Corollary \ref{coro:empty_bins},
\begin{equation*}
B_1\geq \lambda n-2\left(B_0-(1-\lambda) n\right)\in
\left(2-\lambda-2(1+o(1))e^{-\lambda}\right)n
=\frac{2-\lambda-2(1+o(1))e^{-\lambda}}{\lambda}|{\cal C}_v|
\end{equation*}
w.h.p. Since $\lambda\geq 1/4$, the $o(1)$-term is asymptotically negligible.
Without that term, the prefactor is minimized at $\lambda=1$, where it is
strictly larger than $1/4$.

On the other hand, if $\lambda<1/4$, we may w.l.o.g.\ think of the balls as being
thrown sequentially. In this case, the number of balls thrown into occupied
bins is dominated by the sum of $|{\cal C}_v|$ independent Bernoulli variables
taking the value $1$ with probability $1/4$. Since $|{\cal C}_v|\in \omega(\log
n)$, Theorem \ref{theorem:chernoff} yields that w.h.p.\ at most $(1/4+o(1))|{\cal
C}_v|$ balls hit non-empty bins. For sufficiently large $n$, we get that w.h.p.\
more than $(1/2-o(1))|{\cal C}_v|>|{\cal C}_v|/4$ bins receive exactly one ball.

Now assume that instead of being thrown independently, the balls are divided into
$n$ groups of arbitrary size, and the balls from each group are thrown one by one
uniformly at random into the bins that have not been hit by any previous ball of
that group. In this case, the probability to hit an empty bin is always as least
as large as in the previous setting, since only non-empty bins may not be hit by
later balls of a group. Hence, in the end again a fraction larger than
one fourth of the balls are in bins containing no other balls w.h.p.

Finally, consider Step~5 of the algorithm. We identify the copies of messages for
a specific node $v$ with balls and the nodes with bins. The above considerations
show that w.h.p.\ at least $|{\cal C}_v|/4$ nodes receive exactly one of the
copies. Each of these nodes will in Step~6 deliver its copy to the correct
destination. Consider such a node $w\in V$ receiving and forwarding exactly one
message to $v$. Since each node $u\in V$ sends each element of ${\cal C}_v$ with
probability $1/n$ to $w$, the message relayed by $w$ is drawn uniformly at random
from ${\cal C}_v$. Furthermore, as we know that no other copy is sent to $w$, all
other messages are sent with conditional probability $1/(n-1)$ each to any of the
other nodes. Repeating this argument inductively for all nodes receiving exactly
one copy of a message for $v$ in Step~5, we see that the set of messages
transmitted to $v$ by such nodes in Step~6 is a uniformly random subset of
${\cal C}_v$.}

The proof of the main theorem is based on the fact that the number of copies
$\A_s$ creates of each message in Step~5 grows asymptotically exponentially in
each phase as long as it is not too large.
\begin{lemma}\label{lemma:growth_k}
Fix a phase of $\A_s$ and assume that $n$ is sufficiently large. Denote by $m_v$
the number of messages node $v\in V$ still needs to receive and by $k$ the number
of copies of each message created in Step~5 of that phase of $\A_s$. Then, in
Step~6 $v$ will receive at least one copy of all but
$\max\{(1+o(1))e^{-k/4}m_v,e^{-\sqrt{\log n}}n\}$ of these messages w.h.p.
\end{lemma}
\proof{Denote by ${\cal C}_v$ (where $|{\cal C}_v|=km_v$) the set of copies of
messages destined to $v$ that are created in Step~5 of $\A_s$ and assume that
$|{\cal C}_v|\geq e^{-\sqrt{\log n}}n$. Due to Lemma
\ref{lemma:sufficient_messages}, w.h.p.\ a uniformly random subset of size at
least $|{\cal C}_v|/4$ of ${\cal C}_v$ is received by $v$ in Step~6 of that
phase. For each message, exactly $k$ copies are contained in ${\cal C}_v$. Hence,
if we draw elements from ${\cal C}_v$ one by one, each message that we have not
seen yet has probability at least $k/|{\cal C}_v|=1/m_v$ to occur in the next
trial.

Thus, the random experiment where in each step we draw one original message
destined to $v$ u.i.r.\ (with probability $1/m_v$ each) and count the number of
distinct messages stochastically dominates the experiment counting the number of
different messages $v$ receives in Step~6 of the algorithm from below. The former
is exactly the balls-into-bins scenario from Lemma
\ref{lemma:balls_bins_negative}, where (at least) $|{\cal C}_v|/4$ balls are
thrown into $m_v=|{\cal C}_v|/k$ bins. If $k\leq 2\ln |{\cal C}_v|/ \ln \ln n$,
we have
\begin{equation*}
\frac{|{\cal C}_v|}{k\ln \ln n}\ln \left(\frac{|{\cal C}_v|}{k}\right)
\geq \frac{|{\cal C}_v|}{2}\left(1-\frac{\ln k}{\ln|{\cal C}_v|}\right)
\subseteq  \frac{(1-o(1))|{\cal C}_v|}{2}.
\end{equation*}
Hence, Corollary \ref{coro:empty_bins} bounds the number of messages $v$
receives no copy of by
\begin{equation*}
(1+o(1))e^{-k/4}\frac{|{\cal C}_v|}{k}= (1+o(1))e^{-k/4}m_v
\end{equation*}
w.h.p.

On the other hand, if $k$ is larger, we have only a small number of different
messages to deliver. Certainly the bound must deteriorate if we increase this
number at the expense of decreasing $k$ while keeping $|{\cal C}_v|$ fixed (i.e.,
we artificially distinguish between different copies of the same message). Thus,
in this case, we may w.l.o.g.\ assume that $k=\lfloor2\ln |{\cal C}_v|/\ln \ln
n\rfloor\geq 2(\ln n-\sqrt{\log n})/\ln \ln n\gg \sqrt{\log n}$ and apply
Corollary \ref{coro:empty_bins} for this value of $k$, giving that w.h.p.\ $v$
receives all but
\begin{equation*}
(1+o(1))e^{-k/4}\frac{|{\cal C}_v|}{k}\ll e^{-\sqrt{\log n}}n
\end{equation*}
messages.}

We need to show that the algorithm delivers a small number of remaining 
messages quickly.
\begin{lemma}\label{lemma:finalize}
Suppose that $n$ is sufficiently large and fix a phase of $\A_s$. Denote by 
$m_v$ the number of messages node $v\in V$ still needs to receive and by $k$ 
the number of copies created of each message in Step~5. If $m_v\in 
e^{-\Omega(\sqrt{\log n})}n$ and $k\in \Omega(\sqrt{\log n})$, $v$ will 
receive a copy of all messages it has not seen yet within $\BO(1)$ more phases 
of $\A_s$ w.h.p.
\end{lemma}
\proof{Assume w.l.o.g.\ that $m_v k\in \omega(\log n)$; otherwise we simply
increase $m_v$ such that e.g.\ $m_v k\in \Theta((\log n)^2)$ and show that even
when these ``dummy'' messages are added, all messages will be received by $v$
within $\BO(1)$ phases w.h.p.

Thus, according to Lemma \ref{lemma:sufficient_messages}, a uniformly
random fraction of $1/4$ of the copies created for a node $v$ in Step~5 will be
received by it in Step~2. If $k\in \Omega(\log n)$, the probability of a specific
message having no copy in this set is bounded by $(1-1/m_v)^{m_v k/4}\subseteq
e^{-\Omega(\log n)}= n^{-\Omega(1)}$. On the other hand, if $k\in \BO(\log n)$,
each copy has a probability independently bounded from below by $1-km_v/n$ to be
the only one sent to its recipient in Step~5 and thus be delivered successfully
in Step~6. Therefore, for any message the probability of not being delivered in
that phase is bounded by
\begin{eqnarray*}
\left(\frac{km_v}{n}\right)^{k} &\subseteq & 
\left(k e^{-\Omega\left(\sqrt{\log n}\right)}\right)^{k}\\
& \subseteq & e^{-\Omega\left(k\sqrt{\log n}\right)} \\
& \subseteq & e^{-\Omega(\log n)}\\
& = & n^{-\Omega(1)}.
\end{eqnarray*}
Since $m_v$ is non-increasing and $k$ non-decreasing, we conclude that all
messages will be delivered within the next $\BO(1)$ phases w.h.p.}

With this at hand, we can provide a probabilistic upper bound of $\BO(\log^*n)$
on the running time of $\A_s$.
\begin{theorem}\label{theorem:symmetric}
Algorithm $\A_s$ solves Problem~\ref{prob:sidt}. It terminates within
$\BO(\log^* n)$ rounds w.h.p.
\end{theorem}
\proof{Since the algorithm terminates only if all messages have been delivered,
it is correct if it terminates. Since in each phase of $\A_s$ some messages
will reach their destination, it will eventually terminate. A single phase takes
$\BO(1)$ rounds. Hence it remains to show that w.h.p.\ after at most
$\BO(\log^* n)$ phases the termination condition is true, i.e., all messages
have been received at least once by their target nodes.

Denote by $k(i)$ the value $k$ computed in Step~5 of phase $i\in \N$ of $\A_s$, 
by $m_v(i)$ the number of messages a node $v\in V$ still needs to receive in 
that phase, and define $M_r(i):=\max_{v\in V}\{m_v(i)\}$. Thus, we have $k(i)= 
\lfloor n/M_r(i)\rfloor$, where $k(1)=1$. According to Lemma 
\ref{lemma:growth_k}, for all $i$ it holds that w.h.p.\ $M_r(i+1)\in 
\max\{(1+o(1))e^{-k(i)/4}M_r(i),e^{-\sqrt{\log n}}n\}$. Thus, after constantly 
many phases (when the influence of rounding becomes negligible), $k(i)$ starts 
to grow exponentially in each phase, until $M_r(i+1)\leq e^{-\sqrt{\log n}}n$.
According to Lemma \ref{lemma:finalize}, $\A_s$ will w.h.p.\ terminate after
$\BO(1)$ additional phases, and thus after $\BO(\log^* n)$ phases in total.}

\subsection{Tolerance of Transient Link Failures}
Apart from featuring a small running time, $\A_s$ can be adapted in order to
handle substantial message loss and bound the maximum number of duplicates
created of each message. Set $i:=1$ and $k(1):=1$. Given a constant probability
$p\in (0,1)$ of independent link failure, at each node $v\in V$, Algorithm
$\A_l(p)$ executes the following loop until it terminates:
\begin{enumerate}
  \item Create $\lfloor k(i)\rfloor$ copies of each message. Distribute these
  copies uniformly at random among all nodes, but under the constraint that (up
  to one) all nodes receive the same number of messages.
  \item To each node, forward one copy of a message destined to it (if any has
  been received in the previous step; any choice is feasible).
  \item Confirm any received messages.
  \item Forward any confirmations to the original sender of the corresponding
  copy.
  \item Delete all messages for which confirmations have been received and all
  currently held copies of messages.
  \item Set $k(i+1):=\min\{k(i)e^{\lfloor k(i)\rfloor (1-p)^4/5}, \log n\}$
  and $i:=i+1$. If $i>r(p)$, terminate.
\end{enumerate}
Here $r(p)\in \BO(\log^* n)$ is a value sufficiently large to guarantee that
all messages are delivered successfully w.h.p.\ according to
Theorem~\ref{theorem:faults}.

Lemmas~\ref{lemma:sufficient_messages} and Lemma \ref{lemma:growth_k} also
apply to $\A_l(p)$ granted that not too much congestion is created.
\begin{corollary}\label{coro:sufficient_messages_faults}
Denote by ${\cal C}_v$ the set of copies of messages destined to a node $v\in V$
that $\A_l(p)$ generates in Step~1 of a phase. Provided that $n\geq|{\cal
C}_v|\in \omega(\log n)$ and $n$ is sufficiently large, w.h.p.\ the set of
messages $v$ receives in Step~2 contains a uniformly random subset of ${\cal
C}_v$ of size at least $(1-p)^4|{\cal C}_v|/4$.
\end{corollary}
\proof{Due to Theorem \ref{theorem:chernoff}, w.h.p.\ for a subset of
$(1-o(1))(1-p)^4|{\cal C}_v|$ all four consecutive messages in Steps~1 to 4 of
$\A_l(p)$ will not get lost. Since message loss is independent, this is a
uniformly random subset of ${\cal C}_v$. From here the proof proceeds
analogously to Lemma~\ref{lemma:sufficient_messages}.}

\begin{corollary}\label{coro:growth_k_faults}
Fix a phase of $\A_l(p)$ and assume that $n$ is sufficiently large. Denote by
$m_v$ the number of messages node $v\in V$ still needs to receive and by $k$ the
number of copies of each message created in Step~1 of that phase of $\A_l(p)$.
Then $v$ will w.h.p.\ get at least one copy of
all but $\max\{(1+o(1))e^{-(1-p)^4k(i)/4}m_v(i),e^{-\sqrt{\log n}}n\}$ of these
messages.
\end{corollary}
\proof{Analogous to Lemma~\ref{lemma:growth_k}.}

Again we need to show that all remaining messages are delivered quickly once
$k$ is sufficiently large.
\begin{lemma}\label{lemma:finalize_faults}
Suppose that $n$ is sufficiently large and fix a phase of $\A_l(p)$. If $k\in
\Omega(\log n)$, $v$ will w.h.p.\ receive a copy of all messages it has not
seen yet within $\BO(1)$ more phases of $\A_l(p)$.
\end{lemma}
\proof{If $v$ still needs to receive $\omega(1)$ messages, we have that $|{\cal
C}_v|\in \omega(\log n)$. Thus, Corollary~\ref{coro:sufficient_messages_faults}
states that $v$ will receive a uniformly random subset of fraction of at least
$(1-p)^4/4$ of all copies destined to it. Hence, the probability that a specific
message is not contained in this subset is at most
\begin{equation*}
\left(1-\frac{(1-p)^4}{4}\right)^{\Omega(\log n)}\subseteq n^{-\Omega(1)}.
\end{equation*}

On the other hand, if the number of messages for $v$ is small---say, $\BO(\log
n)$---in total no more than $\BO((\log n)^2)$ copies for $v$ will be present in
total. Therefore, each of them will be delivered with probability at least
$(1-o(1))(1-p)^4$ independently of all other random choices, resulting in a
similar bound on the probability that at least one copy of a specific message is
received by $v$. Hence, after $\BO(1)$ rounds, w.h.p.\ $v$ will have received
all remaining messages.}

We now are in the position to bound the running time of $\A_l(p)$.
\begin{theorem}\label{theorem:faults}
Assume that messages are lost u.i.r.\ with probability at most $p<1$, where $p$
is a constant. Then $\A_l(p)$ solves problem \ref{prob:sidt} w.h.p.\ and
terminates within $\BO(\log^*n)$ rounds.
\end{theorem}
\proof{Denote by ${\cal C}_v$ the set of copies destined to $v\in V$ in a phase
of $\A_l(p)$ and by $M_r(i)$ the maximum number of messages any node still needs
to receive and successfully confirm at the beginning of phase $i\in \N$. Provided
that $n\geq |{\cal C}_v|$, Corollary~\ref{coro:growth_k_faults} states that
\begin{equation*}
M_r(i+1)\in \max\left\{(1+o(1))e^{-(1-p)^4\lfloor k(i)\rfloor/4}M_r(i),
e^{-\sqrt{\log n}}n\right\}.
\end{equation*}
The condition that $|{\cal C}_v|\leq n$ is satisfied w.h.p.\ since w.h.p.\ the
maximum number of messages $M_r(i)$ any node must receive in phase $i\in
\{2,\ldots,r(p)\}$ falls faster than $k(i)$ increases and $k(1)=1$. Hence, after
$\BO(\log^* n)$ many rounds, we have $k(i)=\sqrt{\log n}$ and $M_r(i)\leq
e^{-\sqrt{\log n}}n$. Consequently, Corollary \ref{lemma:finalize_faults}
shows that all messages will be delivered after $\BO(1)$ more phases w.h.p.

It remains to show that the number of messages any node still needs to
\emph{send} decreases faster than $k(i)$ increases in order to guarantee that
phases take $\BO(1)$ rounds. Denote by $m_{w\to v}(i)$ the number of messages
node $w\in V$ still needs to send to a node $v\in V$ at the beginning of phase
$i$. Corollary \ref{coro:sufficient_messages_faults} states that for all $v\in V$
w.h.p.\ a uniformly random fraction of at least $(1-p)^4/4$ of the copies
destined to $v$ is received by it. We previously used that the number of
successful messages therefore is stochastically dominated from below by the
number of non-empty bins when throwing $(1-p)^4|C_v|/4$ balls into $|C_v|/k(i)$
bins. Since we are now interested in the number of successful messages from $w$,
we confine this approach to the subset of $m_{w\to v}(i)$ bins corresponding to
messages from $w$ to $v$.\footnote{Certainly a subset of a set of negatively
associated random variables is negatively associated.} Doing the same
computations as in Corollary~\ref{coro:empty_bins}, we see that each message is
delivered with probability at least $1-(1\pm o(1))e^{-(1-p)^4\lfloor
k(i)\rfloor/4}$. Moreover, conditional to the event that all $v\in V$ receive a
uniform subset of $C_v$ of size at least $(1-p)^4|C_v|/4$, these random
experiments are mutually independent for different destinations $v\neq v'\in V$.
Hence we can infer from statement $(ii)$ of Lemma
\ref{lemma:negative_association} that Theorem \ref{theorem:chernoff} is
applicable to the \emph{complete} set of non-empty bins associated with $w$
(i.e., messages from $w$) in these experiments. Thus, analogously to Lemma
\ref{lemma:growth_k}, we conclude that for the total number of messages
$m_w(i):=\sum_{v\in V}m_{w\to v}(i)$ it w.h.p.\ holds that
\begin{equation*}
m_w(i+1)\in \max\left\{(1+o(1))e^{-(1-p)^4\lfloor k(i)\rfloor /4}m_w(i),
e^{-\sqrt{\log n}}n\right\}.
\end{equation*}

This is exactly the same bound as we deduced on the number of messages that still
need to be received in a given phase. Hence, we infer that $m_w(i)$ decreases
sufficiently quickly to ensure that $m_w(i)\lfloor k(i)\rfloor\leq n$ w.h.p.\ and
phases take $\BO(1)$ rounds. Thus, for some appropriately chosen $r(p)\in
\BO(\log^* n)$, after $r(p)$ rounds all nodes may terminate since all messages
have been delivered w.h.p.}

The assumption of independence of link failures can be weakened. E.g.~an u.i.r.\
subset of $p n^2$ links might fail permanently, while all other links are
reliable, i.e., link failures are spatially independent, but temporally fully
dependent.

What is more, similar techniques are applicable to Problems~\ref{prob:idt} and
\ref{prob:bib}. In order to simplify the presentation, we will however return
to the assumption that communication is reliable for the remainder of the paper.
\begin{remark}
Note that it is not possible to devise a terminating algorithm that guarantees
success: If nodes may only terminate when they are certain that all messages
have been delivered, they require confirmations of that fact before they can
terminate. However, to guarantee that all nodes will eventually terminate,
some node must check whether these confirmations arrived at their destinations,
which in turn requires confirmations, and so on. The task reduces to the
(in)famous two generals' problem which is unsolvable.
\end{remark}

\subsection{Solving the General Case}
To tackle Problem \ref{prob:idt}, only a slight modification of $\A_s$ is needed.
At each node $v\in V$, Algorithm~$\A_g$ executes the following loop until
termination:
\begin{enumerate}
  \item Announce the number of currently held messages to all other nodes. If
  no node has any messages left, terminate.
  \item Redistribute the messages evenly such that all nodes store (up to one)
  the same amount.
  \item Announce to each node the number of messages for it you currently hold.
  \item Announce the total number of messages destined for you to all other
  nodes.
  \item Denoting by $M_r'$ the maximum number of messages any node still
  needs to receive, set $k:=\max\{\lfloor n/M_r'\rfloor,1\}$. Create $k$ copies
  of each message and distribute them uniformly at random among all nodes,
  under the constraint that (up to one) over each link the same number of
  messages is sent.
  \item To each node, forward up to $3\lceil M_r'/n\rceil$ copies of messages
  for it (any choices are feasible) and confirm the delivery to the previous
  sender.
  \item Delete all messages for which confirmations have been received and all
  currently held copies of messages.
\end{enumerate}
\begin{theorem}\label{theorem:general}
$\A_g$ solves Problem \ref{prob:idt} w.h.p.\ in
\begin{equation*}
\BO\left(\frac{M_s+M_r}{n}+\left(\log^* n-\log^*
\frac{n}{M_r}\right)\right)
\end{equation*}
rounds.
\end{theorem}
\proof{Denote by $M_r(i)$ the maximum number of messages any node still needs to
receive at the beginning of phase $i\in \N$. The first execution of Step~2 of
the algorithm will take $\lceil M_s/n \rceil$ rounds. Subsequent executions of
Step~2 will take at most $\lceil M_r(i-1)/n \rceil$ rounds, as since the
previous execution of Step~2 the number of messages at each node could not have
increased.

As long as $M_r(i)>n$, each delivered message will be a success since no messages
are duplicated. Denote by $m_v(i)\leq M_r(i)$ the number of messages that still
need to be received by node $v\in V$ at the beginning of the $i^{th}$ phase.
Observe that at most one third of the nodes may get $3m_v(i)/n\leq 3M_r(i)/n$ or
more messages destined to $v$ in Step~5. Suppose a node $w\in V$ holds at most
$2n/3$ messages for $v$. Fix the random choices of all nodes but $w$. Now we
choose the destinations for $w$'s messages one after another uniformly from the
set of nodes that have not been picked by $w$ yet. Until at least $n/3$ many
nodes have been chosen which will certainly deliver the respective messages to
$v$, any message has independent probability of at least $1/3$ to pick such a
node. If $w$ has between $n/3$ and $2n/3$ many messages, we directly apply
Theorem \ref{theorem:chernoff} in order to see that w.h.p.\ a fraction of at
least $1/3-o(1)$ of $w$'s messages will be delivered to $v$ in that phase. For
all nodes with fewer messages, we apply the Theorem to the set subsuming all
those messages, showing that---if these are $\omega(\log n)$ many---again a
fraction of $1/3-o(1)$ will reach $v$. Lastly, any node holding more than $2n/3$
messages destined for $v$ will certainly send more than one third of them to
nodes which will forward them to $v$. Thus, w.h.p.\ $(1/3-o(1))m_v$ messages will
be received by $v$ in that phase granted that $m_v(i)\in \omega(\log n)$.

We conclude that for all phases $i$ we have w.h.p.\ that $M_r(i+1)\leq
\max\left\{n,(2/3+o(1))M_r(i)\right\}$. If in phase $i$ we have $M_r(i)>n$,
Steps~5 and 6 of that phase will require $\BO\left(M_r(i)/n\right)$ rounds.
Hence, Steps~5 and 6 require in total at most
\begin{equation*}
\sum_{i=1}^{\left\lceil\log M_r/n\right\rceil}\BO\left(\frac{M_r(i)}{n}\right)
\subseteq
\BO\left(\frac{M_r}{n}\right)\sum_{i=0}^{\infty}\left(\frac{2}{3}+o(1)\right)^i
\subseteq \BO\left(\frac{M_r}{n}\right)
\end{equation*}
rounds w.h.p.\ until $M_r(i_0)\leq n$ for some phase $i_0$. Taking into account
that Steps~1, 3, 4, and 7 take constant time regardless of the number of
remaining messages, the number of rounds until phase $i_0$ is in
$\BO((M_s+M_r)/n)$ w.h.p.\footnote{To be technically correct, one must mention
that Lemma~\ref{lemma:transitivity} is not applicable if $\log M_r/n$ is not
polynomially bounded in $n$. However, in this extreme case the probability
bounds for failure become much stronger and their sum can be estimated by a
convergent series, still yielding the claim w.h.p.}

In phases $i\geq i_0$, the algorithm will act the same as $\A_s$ did, since
$M_r\leq n$. Thus, as shown for Theorem~\ref{theorem:symmetric}, $k(i)$ will grow
asymptotically exponentially in each step. However, if we have few messages right
from the beginning, i.e., $M_r\in o(n)$, $k(i_0)=k(1)=\lfloor n/M_r\rfloor$
might already be large. Starting from that value, it requires merely $\BO(\log^*
n - \log^* (n/M_r))$ rounds until the algorithm terminates w.h.p. Summing the
bounds on the running time until and after phase $i_0$, the time complexity
stated by the theorem follows.}

Roughly speaking, the general problem can be solved with only constant-factor
overhead unless $M_r\approx n$ and not $M_s\gg n$, i.e., the parameters are close
to the special case of Problem~\ref{prob:sidt}. In this case the solution is
slightly suboptimal. Using another algorithm, time complexity can be kept
asymptotically optimal. Note, however, that the respective algorithm is more
complicated and---since $\log^* n$ grows extremely slowly---will in practice
always exhibit a larger running time.
\begin{theorem}\label{theorem:optimal}
An asymptotically optimal randomized solution of Problem~\ref{prob:sidt} exists.
\end{theorem}
We postpone the proof of this theorem until later, as it relies on a
balls-into-bins algorithm presented in Section~\ref{sec:upper}.

\section{Relation to Balls-into-Bins}\label{sec:balls_bins}
The proofs of Theorems~\ref{theorem:symmetric},~\ref{theorem:faults} and
\ref{theorem:general} repeatedly refer to the classical experiment of throwing
$M$ balls u.i.r.\ into $N$ bins. Indeed, solving Problems~\ref{prob:idt} can be
seen as solving $n$ balls-into-bins problems in parallel, where the messages
for a specific destination are the balls and the relaying nodes are the bins.
Note that the fact that the ``balls'' are not anonymous does not simplify the
task, as labeling balls randomly by $\BO(\log n)$ bits guarantees w.h.p.\
globally unique identifiers for ``anonymous'' balls.

In this section we will show that our technique yields strong bounds for the
well-known distributed balls-into-bins problem formulated by
Adler~et~al.~\cite{adler95}. Compared to their model, the decisive difference is
that we drop the condition of non-adaptivity, i.e., balls do not have to choose a
fixed number of bins to communicate with right from the start.

\subsection{Model}
The system consists of $n$ bins and $n$ balls, and we assume it to be
fault-free. We employ a synchronous message passing model, where one round
consists of the following steps:
\begin{enumerate}
  \item Balls perform (finite, but otherwise unrestricted) local computations
  and send messages to arbitrary bins.
  \item Bins receive these messages, do local computations, and send messages to
  any balls they have been contacted by in this or earlier rounds.
  \item Balls receive these messages and may commit to a bin (and
  terminate).\footnote{Note that (for reasonable algorithms) this step does not
  interfere with the other two. Hence, the literature typically accounts for
  this step as ``half a round'' when stating the time complexity of
  balls-into-bins algorithms; we adopted this convention in the related work
  section.}
\end{enumerate}
Moreover, balls and bins each have access to an unlimited source of unbiased
random bits, i.e., all algorithms are randomized. The considered task now can be
stated concisely.
\begin{problem}[Parallel Balls-into-Bins]\label{prob:bib}
We want to place each ball into a bin. The goals are to minimize the total
number of rounds until all balls are placed, the maximum number of balls placed
into a bin, and the amount of involved communication.
\end{problem}

\subsection{Basic Algorithm}
Algorithm $\A_s$ solved Problem \ref{prob:sidt} essentially by partitioning it
into $n$ balls-into-bins problems and handling them in parallel. We extract the
respective balls-into-bins algorithm.

Set $k(1):=1$ and $i=1$. Algorithm $\A_b$ executes the following loop until
termination:
\begin{enumerate}
  \item Balls contact $\lfloor k(i)\rfloor$ u.i.r.\ bins, requesting permission
  to be placed into them.
  \item Each bin admits permission to one of the requesting balls (if any) and
  declines all other requests.
  \item Any ball receiving at least one permission chooses an arbitrary of
  the respective bins to be placed into, informs it, and terminates.
  \item Set $k(i+1):=\min\{k(i)e^{\lfloor k(i)\rfloor/5}, \sqrt{\log n}\}$ and
  $i:=i+1$.
\end{enumerate}
\begin{theorem}\label{theorem:bib}
$\A_b$ solves Problem \ref{prob:bib}, guaranteeing the following properties:
\begin{itemize}
  \item It terminates after $\log^* n + \BO(1)$ rounds w.h.p.
  \item Each bin in the end contains at most $\log^* n+\BO(1)$ balls w.h.p.
  \item In each round, the total number of messages sent is w.h.p.\ at most
  $n$. The total number of messages is w.h.p.\ in $\BO(n)$.
  \item Balls send and receive $\BO(1)$ messages in expectation and
  $\BO(\sqrt{\log n})$ many w.h.p.
  \item Bins send and receive $\BO(1)$ messages in expectation and $\BO(\log
  n/\log \log n)$ many w.h.p.
\end{itemize}
Furthermore, the algorithm runs asynchronously in the sense that balls and bins
can decide on any request respectively permission immediately, provided that
balls' messages may contain round counters. According to the previous statements
messages then have a size of $\BO(1)$ in expectation and $\BO(\log \log^* n)$
w.h.p.
\end{theorem}
\proof{The first two statements and the first part of the third follow as
corollary of Theorem~\ref{theorem:symmetric}. Since it takes only one round to
query a bin, receive its response, and decide on a bin, the time complexity
equals the number of rounds until $k(i)=\sqrt{\log n}$ plus $\BO(1)$ additional
rounds. After $\BO(1)$ rounds, we have that $k(i)\geq 20$ and thus $k(i+1)\geq
\min\{k(i)e^{k(i)/20},\sqrt{\log n}\}$, i.e., $k(i)$ grows at least like an
exponential tower with basis $e^{1/20}$ until it reaches $\sqrt{\log n}$. As we
will verify in Lemma~\ref{lemma:calc}, this takes $\log^* n + \BO(1)$ steps. The
growth of $k(i)$ and the fact that algorithm terminates w.h.p.\ within $\BO(1)$
phases once $k(i)=\sqrt{\log n}$ many requests are sent by each ball implies that
the number of messages a ball sends is w.h.p.\ bounded by $\BO(\sqrt{\log n})$.

Denote by $b(i)$ the number of balls that do not terminate w.h.p.\ until the
beginning of phase $i\in \N$. Analogously to Theorem \ref{theorem:general} we
have w.h.p., so in particular with probability $p\geq 1-1/n^2$, that
\begin{equation}\label{eq:shrinking}
b(i+1)\leq (1+o(1))\max\left\{b(i)e^{-\lfloor k(i)\rfloor /4},
e^{-\sqrt{\log n}}n\right\}
\end{equation}
for all $i\in \{1,\ldots,r\in \BO(\log^* n)\}$ and the algorithm terminates
within $r$ phases. In the unlikely case that these bounds do not hold, certainly
at least one ball will be accepted in each round. Let $i_0\in \BO(1)$ be the
first phase in which we have $k(i_0)\geq 40$. Since each ball follows the same
strategy, the random variable $X$ counting the number of requests it sends does
not depend on the considered ball. Thus we can bound
\begin{eqnarray*}
E(X)&< & \frac{1}{n}\left((1-p)\sum_{i=1}^n i\sqrt{\log n}
+p\left(\sum_{i=1}^r b(i)\lfloor k(i)\rfloor\right)\right)\\
&\in & o(1)+\frac{1}{n}\left(\sum_{i=1}^{i_0}b(i)k(i)
+\sum_{i=i_0+1}^r b(i) k(i)\right)\\
&\subseteq & \BO(1)+\frac{1}{n}\left(\sum_{i=i_0+1}^r b(i_0)k(i_0)
\left(\max_{j\geq i_0} 
\left\{e^{-\lfloor k(j)\rfloor/4+\lfloor k(j)\rfloor /5}\right\}\right)^{i-i_0}
+e^{-\sqrt{\log n}}n\right)\\
&\subseteq & \BO(1)+\frac{1}{n}\sum_{i=i_0+1}^r b(i_0)k(i_0)
 e^{-(i-i_0)(\lfloor k(i_0)/20 \rfloor-1)}\\
&\subseteq & \BO(1)+\frac{b(i_0)k(i_0)}{n}\sum_{i=1}^{\infty}e^{-i}\\
&\subseteq & \BO(1).
\end{eqnarray*}
Thus, each ball sends in expectation $\BO(1)$ messages. The same is true for
bins, as they only answer to balls' requests. Moreover, since we computed that
\begin{equation*}
\sum_{i=1}^r b(i)\lfloor k(i)\rfloor\in \BO(n)
\end{equation*}
w.h.p., at most $\BO(n)$ messages are sent in total w.h.p.

Next, we need to show the upper bound on the number of messages bins receive (and
send) w.h.p. To this end, assume that balls send $4c m\in 2\N$ messages to
u.i.r.\ bins in rounds when they would send $m\leq \sqrt{\log n}$ messages
according to the algorithm. The probability for any such message to be received
by the same destination as another is independently bounded by $4c m/n$.
Since $\binom{n}{k}\leq (en/k)^k$, the probability that this happens for any
subset of $2c m$ messages is less than
\begin{equation*}
\binom{4c m}{2c m}\left(\frac{4c m}{n}\right)^{2c m}\leq 
\left(\frac{8c e m}{n}\right)^{2c m}< n^{-c},
\end{equation*}
i.e., w.h.p.\ at least $2c m\geq m$ different destinations receive a message.

Thus, we may w.l.o.g.\ bound the number of messages received by the bins in this
simplified scenario instead. This is again the balls-into-bins experiment from
Lemma \ref{lemma:balls_bins_negative}. Because at most $\BO(n)$ messages
are sent in total w.h.p., Theorem~\ref{theorem:chernoff} shows that w.h.p.\ at
most $\BO(\log n/\log \log n)$ messages are received by each bin as claimed.

Finally, the proof is not affected by the fact that bins admit the first ball
they receive a message with a given round counter from if communication is
asynchronous, since Corollary \ref{coro:sufficient_messages_faults} only
considers nodes (i.e., bins) receiving exactly one request in a given round.
Hence all results directly transfer to the asynchronous case.}
\begin{remark}
Note that the amount of communication caused by Algorithms $\A_s$, $\A_l(p)$,
and $\A_g$ can be controlled similarly. Moreover, these algorithms can be
adapted for asynchronous execution as well.
\end{remark}

\subsection{Variations}
Our approach is quite flexible. For instance, we can ensure a bin load of at
most two without increasing the time complexity.
\begin{corollary}\label{coro:constant}
We modify $\A_b$ into $\A_b^2$ by ruling that any bins having already accepted
two balls refuse any further requests in Step~2, and in Step~4 we set 
\begin{equation*}
k(i+1):=\min\{k(i)e^{\lfloor k(i)\rfloor/10}, \log n\}.
\end{equation*}
Then the statements of Theorem~\ref{theorem:bib} remain true except that balls
now send w.h.p.\ $\BO(\log n)$ messages instead of $\BO(\sqrt{\log n})$. In
turn, the maximum bin load of the algorithm becomes two.
\end{corollary}
\proof{As mentioned before, Theorem~\ref{theorem:faults} also applies if we fix
a subset of the links that may fail. Instead of failing links, we now have
``failing bins'', i.e., up to one half of the bins may reject any requests
since they already contain two balls. This resembles a probability of $1/2$
that a ball is rejected despite it should be accepted, i.e., the term of
$(1-p)^4$ in Algorithm $\A_l(p)$ is replaced by $1/2$.

Having this observation in mind, we can proceed as in the proof of
Theorem~\ref{theorem:bib}, where Theorem~\ref{theorem:faults} takes the role of
Theorem~\ref{theorem:symmetric}.}

If we start with less balls, the algorithm terminates quickly.
\begin{corollary}\label{coro:quick}
If only $m:=n/\log^{(r)}n$ balls are to be placed into $n$ bins for some
$r\in \N$, $\A_b^2$ initialized with $k(1):=\lfloor \log^{(r)}n \rfloor$
terminates w.h.p.\ within $r+\BO(1)$ rounds.
\end{corollary}
\proof{This can be viewed as the algorithm being started in a later round, and
only $\log^*n - \log^* (\log^{(r)} n)+\BO(1)=r+\BO(1)$ more rounds are required
for the algorithm to terminate.}

What is more, if a constant time complexity is in demand, we can enforce it at
the expense of an increase in maximum bin load.
\begin{corollary}\label{coro:algo_constant}
For any $r\in \N$, $\A_b$ can be modified into an Algorithm $\A_b(r)$
that guarantees a maximum bin load of $\log^{(r)}n/\log^{(r+1)}n+r+\BO(1)$
w.h.p.\ and terminates within $r+\BO(1)$ rounds w.h.p. Its message complexity
respects the same bounds as the one of $\A_b$.
\end{corollary}
\proof{In order to speed up the process, we rule that in the first phase bins
accept up to $l:=\lfloor\log^{(r)}n/\log^{(r+1)}n\rfloor$ many balls. Let for
$i\in \N_0$ $Y^i$ denote the random variables counting the number of bins with at
least $i$ balls in that phase. From Lemma \ref{lemma:balls_bins_negative} we know
that Theorem \ref{theorem:chernoff} applies to these variables, i.e., $Y^i\in
\BO(E(Y^i)+\log n)$ w.h.p. Consequently, the same is true for the number
$Y^i-Y^{i+1}$ of bins receiving \emph{exactly} $i$ messages. Moreover, we already
observed that w.h.p.\ bins receive $\BO(\log n/\log \log n)$ messages. Thus, the
number of balls that are not accepted in the first phase is w.h.p.\ bounded by
\begin{eqnarray*}
\polylog n+\BO\left(n\sum_{i=l+1}^{n}(i-l)\binom{n}{i}
\left(\frac{1}{n}\right)^i\left(1-\frac{1}{n}\right)^{n-i}\right)
&\subseteq &\polylog n+\BO\left(n\sum_{i=l}^{\infty}
\left(\frac{e}{i}\right)^i\right)\\
&\subseteq &\polylog n+n\sum_{i=l}^{\infty}\Omega(l)^{-i}\\
&\subseteq &\polylog n+2^{-\Omega(l\log l)}n,
\end{eqnarray*}
where in the first step we used the inequality $\binom{n}{k}\leq (en/k)^k$.

Thus, after the initial phase, w.h.p.\ only $n/(\log^{(r-1)}n)^{\Omega(1)}$ 
balls remain. Hence, in the next phase, $\A_b(r)$ may proceed as $\A_b$, but 
with $k(2)\in (\log^{(r-1)}n)^{\Omega(1)}$ requests per ball; we conclude that 
the algorithm terminates within $r +\BO(1)$ additional rounds w.h.p.}

The observation that neither balls nor bins need to wait prior to deciding on
any message implies that our algorithms can also be executed sequentially,
placing one ball after another. In particular, we can guarantee a bin load of
two efficiently. This corresponds to the simple sequential algorithm that queries
for each ball sufficiently many bins to find one that has load less than two.
\begin{lemma}\label{lemma:sequential}
An adaptive sequential balls-into-bins algorithm $\A_{\mbox{seq}}$ exists
guaranteeing a maximum bin load of two, requiring at most $(2+o(1))n$
random choices and bin queries w.h.p.
\end{lemma}
\proof{The algorithm simply queries u.i.r.\ bins until one of load less than two
is found; then the current ball is placed and the algorithm proceeds with the
next. Since at least half of the bins have load less than two at any time, each
query has independent probability of $1/2$ of being successful. Therefore, it can
be deduced from Theorem~\ref{theorem:chernoff}, that w.h.p.\ no more than
$(2+o(1))n$ bin queries are necessary to place all balls.}

\section{Lower Bound}\label{sec:lower}
In this section, we will derive our lower bound on the parallel complexity of the
balls-into-bins problem. After presenting the formal model and initial
definitions, we proceed by proving the main result. Subsequently, we briefly
present some generalizations of our technique.

\subsection{Definitions}

A natural restriction for algorithms solving Problem~\ref{prob:bib} is to assume
that random choices cannot be biased, i.e., also bins are anonymous. This is
formalized by the following definition.
\begin{problem}[Symmetric Balls-into-Bins]\label{prob:sbib}
We call an instance of Problem~\ref{prob:bib} \emph{symmetric parallel
balls-into-bins problem}, if balls and bins identify each other by u.i.r.\ port
numberings. We call an algorithm solving this problem \emph{symmetric}.
\end{problem}
Thus, whenever a ball executing a symmetric balls-into-bins algorithm contacts a
new bin, it essentially draws uniformly at random. This is a formalization of the
central aspect of the notion of symmetry used by Adler et al.~\cite{adler95}.

Recall that the symmetric Algorithm~$\A_b^2$ solves Problem~\ref{prob:bib} in
$\log^* n+\BO(1)$ rounds with a maximum bin load of two, using w.h.p.\ $\BO(n)$
messages in total. We will prove that the time complexity of symmetric
algorithms cannot be improved by any constant factor, unless considerably more
communication is used or larger bin loads are tolerated. Moreover, our lower
bound holds for a stronger communication model.
\begin{problem}[Acquaintance Balls-into-Bins]\label{prob:acbib}
We call an instance of Problem~\ref{prob:bib} \emph{acquaintance balls-into-bins
problem}, if the following holds. Initially, bins are anonymous, i.e., balls
identify bins by u.i.r.\ port numberings. However, once a ball contacts a bin, it
learns its globally unique address, by which it can be contacted reliably. Thus,
by means of forwarding addresses, balls can learn to contact specific bins
directly. The addresses are abstract in the sense that they can be used for
this purpose only.\footnote{This requirement is introduced to permit the use of
these addresses for symmetry breaking, as is possible for asymmetric algorithms.
One may think of the addresses e.g.\ as being random from a large universe, or
the address space might be entirely unknown to the balls.} We call an
algorithm solving this problem \emph{acquaintance algorithm}.
\end{problem}

We will show that any acquaintance algorithm guaranteeing w.h.p.\ $\BO(n)$ total
messages and $\polylog n$ messages per node requires w.h.p.\ at least
$(1-o(1))\log^* n$ rounds to achieve a maximum bin load of $^{o(\log^*
n)}2$.\footnote{By $^ka$ we denote the tetration, i.e., $k$ times iterated
exponentiation by $a$.}

We need to bound the amount of information balls can collect during the
course of the algorithm. As balls may contact any bins they heard of, this is
described by exponentially growing neighborhoods in the graph where edges are
created whenever a ball picks a communication partner at random.
\begin{definition}[Balls-into-Bins Graph]\label{def:bib_graph}
The (bipartite and simple) \emph{balls-into-bins graph} $G_{\A}(t)$ associated
with an execution of the acquaintance algorithm $\A$ running for $t\in \N$ rounds
is constructed as follows. The node set $V:=V_{\circ}\cup V_{\sqcup}$ consists of
$|V_{\circ}|=|V_{\sqcup}|=n$ bins and balls. In each round $i\in \{1,\ldots,t\}$,
each ball $b\in V_{\circ}$ adds an edge connecting itself to bin $v\in
V_{\sqcup}$ if $b$ contacts $v$ by a random choice in that round. By $E_{\A}(i)$
we denote the edges added in round $i$ and $G_{\A}(t)=(V,\cup_{i=1}^t
E_{\A}(i))$ is the graph containing all edges added until and including
round~$t$.
\end{definition}
In the remainder of the section, we will consider such graphs only.

The proof will argue about certain symmetric subgraphs in which not all balls
can decide on bins concurrently without incurring large bin loads. As can be
seen by a quick calculation, any connected subgraph containing a cycle is
unlikely to occur frequently. For an adaptive algorithm, it is possible that
balls make a larger effort in terms of sent messages to break symmetry once
they observe a ``rare'' neighborhood. Therefore, it is mandatory to reason
about subgraphs which are trees.

We would like to argue that any algorithm suffers from generating a large number
of trees of uniform ball and bin degrees. If we root such a tree at an arbitrary
bin, balls cannot distinguish between their parents and children according to
this orientation. Thus, they will decide on a bin that is closer to the root with
probability inverse proportional to their degree. If bin degrees are by factor
$f(n)$ larger than ball degrees, this will result in an expected bin load of the
root of $f(n)$. However, this line of reasoning is too simple. As edges are added
to $G$ in different rounds, these edges can be distinguished by the balls.
Moreover, even if several balls observe the same local topology in a given round,
they may randomize the number of bins they contact during that round, destroying
the uniformity of degrees. For these reasons, we $(i)$ rely on a more complicated
tree in which the degrees are a function of the round number and $(ii)$ show that
for every acquaintance algorithm a stronger algorithm exists that indeed
generates many such trees w.h.p.

In summary, the proof will consist of three steps. First, for any acquaintance
algorithm obeying the above bounds on running time and message complexity, an
equally powerful algorithm from a certain subclass of algorithms exists. Second,
algorithms from this subclass w.h.p.\ generate for $(1-o(1))\log^* n$ rounds
large numbers of certain highly symmetric subgraphs in $G_{\A}(t)$. Third,
enforcing a decision from all balls in such structures w.h.p.\ leads to a maximum
bin load of $\omega(1)$.

The following definition clarifies what we understand by ``equally powerful'' in
this context.
\begin{definition}[W.h.p.\ Equivalent Algorithms]
We call two Algorithms $\A$ and $\A'$ for Problem~\ref{prob:bib} \emph{w.h.p.\
equivalent} if their output distributions agree on all but a fraction of
the events occurring with total probability at most $1/n^c$, where $c>0$ is a
tunable constant. That is, if $\Gamma$ denotes the set of possible
distributions of balls into bins, we have that
\begin{equation*}
\sum_{\gamma \in \Gamma}|P_{\A}(\gamma)-P_{\A'}(\gamma)|\leq \frac{1}{n^c}. 
\end{equation*}
\end{definition}

The subclass of algorithms we are interested in is partially characterized by its
behaviour on the mentioned subgraphs, hence we need to define the latter first.
These subgraphs are special trees, in which all involved balls up to a certain
distance from the root see exactly the same topology. This means that $(i)$ in
each round, all involved balls created exactly the same number of edges by
contacting bins randomly, $(ii)$ each bin has a degree that depends on the round
when it was contacted first only, $(iii)$ all edges of such bin are formed in
exactly this round, and $(iv)$ this scheme repeats itself up to a distance that
is sufficiently large for the balls not to see any irregularities that might help
in breaking symmetry. These properties are satisfied by the following recursively
defined tree structure.

\begin{figure}
\begin{center}
\includegraphics[width=\textwidth]{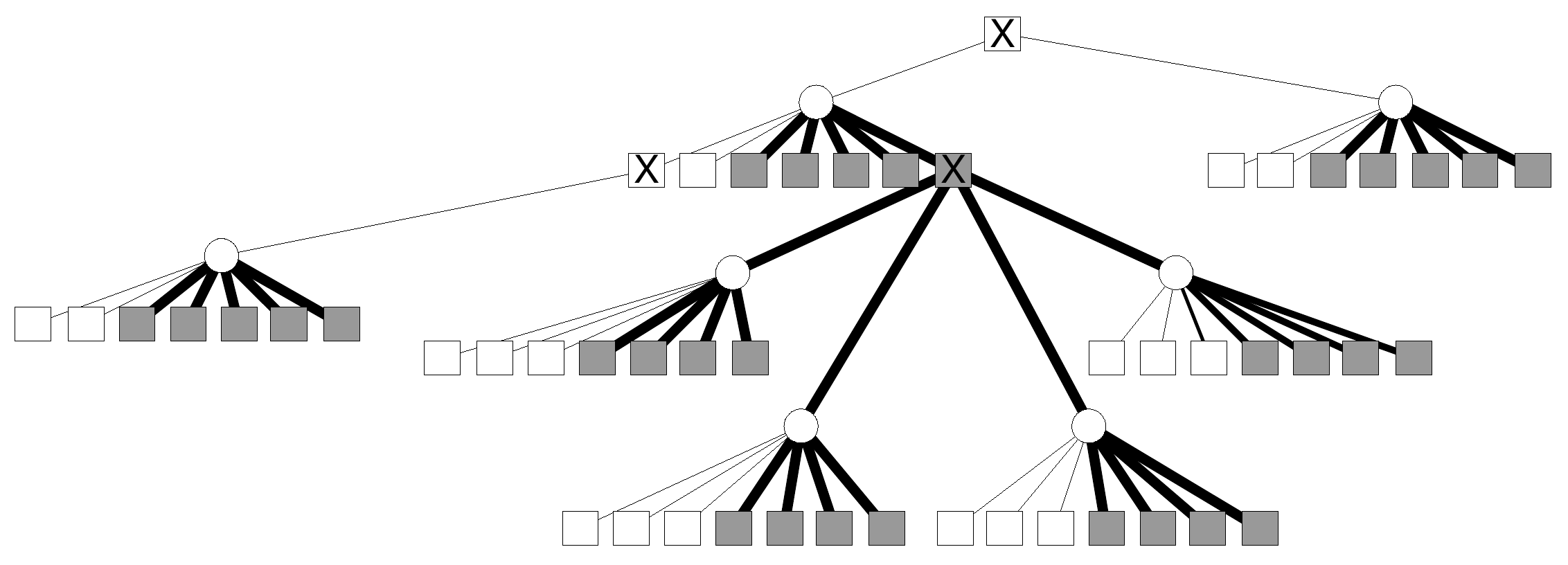}
\caption{Part of a $((2,5),(3,5),D)$-tree rooted at the topmost bin. Bins are
squares and balls are circles; neighborhoods of all balls and the bins marked by
an ``X'' are depicted completely, the remainder of the tree is left out. Thin
edges and white bins were added to the structure in the first round, thick
edges and grey bins in the second. Up to distance $2D$ from the root, the
pattern repeats itself, i.e., the $(2D-d)$-neighborhoods of all balls up to
depth $d$ appear identical.}\label{fig:build_tree}
\end{center}
\end{figure}

\begin{definition}[Layered $(\Dbi,\Dba,D)$-Trees]
A \emph{layered $(\Dbi,\Dba,D)$-tree} of $\ell\in \N_0$ \emph{levels}
rooted at bin $R$ is defined as follows, where
$\Dbi=(\Dbi_1,\ldots,\Dbi_{\ell})$ and $\Dba=(\Dba_1,\ldots,\Dba_{\ell})$ are
the vectors of bins' and balls' degrees on different levels, respectively.

If $\ell=0$, the ``tree'' is simply a single bin. If $\ell>0$, the subgraph
of $G_{\A}(\ell)$ induced by ${\cal N}_R^{(2D)}$ is a tree, where ball degrees
are uniformly $\sum_{i=1}^{\ell}\Dba_i$. Except for leaves, a bin that is added
to the structure in round $i\in \{1,\ldots,\ell\}$ has degree $\Dbi_i$ with all
its edges in $E_{\A}(i)$. See Figure~\ref{fig:build_tree} for an illustration.
\end{definition}

Intuitively, layered trees are crafted to present symmetric neighborhoods
to nodes which are not aware of leaves. Hence, if bins' degrees are large
compared to balls' degrees, not all balls can decide simultaneously without
risking to overload bins. This statement is made mathematically precise later.

We are now in the position to define the subclass of algorithms we will analyze.
The main reason to resort to this subclass is that acquaintance algorithms may
enforce seemingly asymmetric structures, which complicates proving a lower bound.
In order to avoid this, we grant the algorithms additional random choices,
restoring symmetry. The new algorithms must be even stronger, since they have
more information available, yet they will generate many layered trees. Since we
consider such algorithms specifically for this purpose, this is hard-wired in
the definition.

\begin{definition}[Oblivious-Choice Algorithms]\label{def:oblivious_choice}
Assume that given $\Dbi=(\Dbi_1,\ldots,\Dbi_t)$, $\Dba=(\Dba_1,\ldots,\Dba_t)$,
and an acquaintance Algorithm $\A$, we have a sequence $T=(T_0,\ldots,T_t)$,
such that $T_i$ lower bounds w.h.p.\ the number of disjoint layered
$((\Dbi_1,\ldots,\Dbi_i),(\Dba_1,\ldots,\Dba_i),2^t)$-trees in $G_{\A}(i)$
and for all $i\in\{1,\ldots,t\}$ it holds that $\Dba_i\in \BO(n/T_{i-1})$.

We call $\A$ a $(\Dbi,\Dba,T)$-\emph{oblivious-choice algorithm}, if the
following requirements are met:
\begin{itemize}
  \item[(i)] The algorithm terminates at the end of round $t$, when all
  balls simultaneously decide into which bin they are placed. A ball's decision
  is based on its $2^t$-neighborhood in $G_{\A}(t)$, including the random bits
  of any node within that distance, and all bins within this distance are
  feasible choices.\footnote{This is a superset of the information a ball can
  get when executing an acquaintance algorithm, since by address forwarding it
  might learn of and contact bins up to that distance. Note that randomly
  deciding on an unknown bin here counts as contacting it, as a single round
  makes no difference with respect to the stated lower bound.}
  \item[(ii)] In round $i$, each ball $b$ decides on a number of bins to contact
  and chooses that many bins u.i.r., forming the respective edges in
  $G_{\A}(i)$ if not yet present. This decision may resort to the topology of
  the $2^t$-hop neighborhood of a ball in $G_{\A}(i-1)$ (where $G_{\A}(0)$
  is the graph containing no edges).
  \item[(iii)] In round $i$, it holds w.h.p.\ for $\Omega(T_{i-1})$ layered
  $((\Dbi_1,\ldots,\Dbi_{i-1}),(\Dba_1,\ldots,\Dba_{i-1}),2^t)$-trees in
  $G_{\A}(i)$ that all balls in depth $d\leq 2^t$ of such a tree choose
  $\Dba_i$ bins to contact.
\end{itemize}
\end{definition}

The larger $t$ can be, the longer it will take until eventually no more layered
trees occur and all balls may decide safely.

\subsection{Proof}
We need to show that for appropriate choices of parameters and non-trivial values
of $t$, indeed oblivious-choice algorithms exist. Essentially, this is a
consequence of the fact that we construct trees: When growing a tree, each added
edge connects to a node outside the tree, therefore leaving a large number of
possible endpoints of the edge; in contrast, closing a circle in a small subgraph
is unlikely.

\begin{lemma}\label{lemma:sufficient_trees}
Let $\Dba_1\in \N$ and $C>0$ be constants, $L, t\in \N$ arbitrary,
$T_0:=n/(100(\Dba_1)^2(2C+1)L)\in \Theta(n/L)$, and $\Dbi_1:=2L\Dba_1$.
Define for $i\in \{2,\ldots,t\}$ that
\begin{eqnarray*}
\Dba_i &:=& \left\lceil\frac{\Dba_1n}{T_{i-1}}\right\rceil,\\
\Dbi_i &:=& 2L\Dba_i,
\end{eqnarray*}
and for $i\in \{1,\ldots,t\}$ that
\begin{equation*}
T_i := 2^{-(n/T_{i-1})^{4\cdot 2^t}}n.
\end{equation*}
If $T_t\in \omega(\sqrt{n}\log n)$ and $n$ is sufficiently large, then any
algorithm fulfilling the prerequisites $(i)$, $(ii)$, and $(iii)$ from
Definition~\ref{def:oblivious_choice} with regard to these parameters that
sends at most $C n^2/T_{i-1}$ messages in round $i\in \{1,\ldots,t\}$ w.h.p.\ is
a $(\Dbi,\Dba,T)$-oblivious-choice algorithm.
\end{lemma}
\proof{Since by definition we have $\Dba_i\in \BO(n/T_{i-1})$ for all $i\in
\{1,\ldots,t\}$, in order to prove the claim we need to show that at least $T_i$
disjoint layered $((\Dbi_1,\ldots,\Dbi_i),(\Dba_1,\ldots,\Dba_i),2^t)$-trees
occur in $G_{\A}(i)$ w.h.p. We prove this statement by induction. Since $T_0\leq
n$ and every bin is a $((),(),2^t)$-tree, we need to perform the induction step
only.

Hence, assume that for $i-1\in \{0,\ldots,t-1\}$, $T_{i-1}$ lower bounds the
number of disjoint layered
$((\Dbi_1,\ldots,\Dbi_{i-1}),(\Dba_1,\ldots,\Dba_{i-1}),2^t)$-trees in
$G_{\A}(i-1)$ w.h.p. In other words, the event ${\cal E}_1$ that we have at least
$T_{i-1}$ such trees occurs w.h.p.

We want to lower bound the probability $p$ that a so far isolated bin $R$ becomes
the root of a $((\Dbi_1,\ldots,\Dbi_i),(\Dba_1,\ldots,\Dba_i),2^t)$-tree in
$G_{\A}(i)$. Starting from $R$, we construct the $2D$-neighborhood of $R$. All
involved balls take part in disjoint
$((\Dbi_1,\ldots,\Dbi_{i-1}),(\Dba_1,\ldots,\Dba_{i-1}),2^t)$-trees, all bins
incorporated in these trees are not adjacent to edges in $E_{\A}(i)$, and all
bins with edges on level $i$ have been isolated until and including round $i-1$.

As the algorithm sends at most $\sum_{j=1}^{i-1}C n^2/T_{j-1}$ messages until the
end of round $i-1$ w.h.p., the expected number of isolated bins after round
$i-1$ is at least
\begin{eqnarray*}
\left(1-\frac{1}{n^c}\right)n\left(1-\frac{1}{n}\right)^{C n
\sum_{j=1}^{i-1}n/T_{j-1}}&\in & n e^{-(1+o(1))Cn/T_{i-1}}\\
&\subset & n e^{-\BO(n/T_{t-1})}\\
&\subset & \omega(\log n).
\end{eqnarray*}
Thus Lemma~\ref{lemma:balls_bins_negative} and Theorem~\ref{theorem:chernoff}
imply that the event ${\cal E}_2$ that at least $n e^{-(1+o(1))Cn/T_{i-1}}$
such bins are available occurs w.h.p.

Denote by $N$ the total number of nodes in the layered tree. Adding balls one by
one, in each step we choose a ball out of w.h.p.\ at least $T_{i-1}-N+1$
remaining balls in disjoint
$((\Dbi_1,\ldots,\Dbi_{i-1}),(\Dba_1,\ldots,\Dba_{i-1}),2^t)$-trees, connect it
to a bin already in the tree, and connect it to $\Dba_i-1$ of the w.h.p.\ at
least $n e^{-(1+o(1))C n/T_{i-1}}-N+1$ remaining bins that have degree zero in
$G_{\A}(i-1)$. Denote by ${\cal E}_3$ the event that the tree is constructed
successfully and let us bound its probability.

Observe that because for all $i\in \{1,\ldots,t\}$ we have that
$\Dbi_i>2\Dbi_{i-1}$ and $\Dba_i>2\Dba_{i-1}$, it holds that
\begin{equation}
N<\sum_{d=0}^{2^t}\left(\Dbi_i\left(\sum_{j=1}^i\Dba_j\right)\right)^d
<\sum_{d=0}^{2^t}\left(2\Dbi_i\Dba_i\right)^d
<2\left(2\Dbi_i\Dba_i\right)^{2^t}.\label{eq:N_strong}
\end{equation}
Furthermore, the inductive definitions of $\Dbi_i$, $\Dba_i$, and $T_i$, the
prerequisite that $T_t\in \omega(\sqrt{n}\log n)$, and basic calculations reveal
that for all $i\in \{1,\ldots,t\}$, we have the simpler bound of
\begin{equation}
N < 2\left(2\Dbi_i\Dba_i\right)^{2^t}< 2(4L+1)^{2t}\left(\frac{\Dbi_1
n}{T_{i-1}}\right)^{4t}\in ne^{-\omega(n/T_{i-1})}\cap o(T_{i-1})
\label{eq:N_weak}
\end{equation}
on $N$.

 Thus, provided that ${\cal
E}_1$ occurs, the (conditional) probability that a bin that has already been
attached to its parent in the tree is contacted by the first random choice of
exactly $\Dbi_i-1$ balls that are sufficiently close to the roots of disjoint
$((\Dbi_1,\ldots,\Dbi_{i-1}),(\Dba_1,\ldots,\Dba_{i-1}),2^t)$-trees is lower
bounded by
\begin{eqnarray*}
&&\binom{T_{i-1}-N+(\Dbi_i-1)}{\Dbi_i-1}
\left(\frac{1}{n}\right)^{\Dbi_i-1}
\left(1-\frac{1}{n}\right)^{(\Dbi_i-1)(\Dba_i-1)}\\
&\sr{eq:N_weak}{\in}& \left(\frac{T_{i-1}}{n\Dbi_i}\right)^{(1+o(1))(\Dbi_i-1)}.
\end{eqnarray*}
Because $\Dbi_i\in \BO(n/T_{i-1})$, it holds that $\ln(n\Dbi_i/T_{i-1})\in
o(n/T_{i-1})$. Thus, going over all bins (including the root, where the
factor in the exponent is $\Dbi_i$ instead of $\Dbi_i-1$), we can lower bound
the probability that all bins are contacted by the right number of balls by
\begin{equation*}
\left(\frac{T_{i-1}}{n\Dbi_i}\right)^{(1+o(1))N}\in e^{-(1+o(1))N n/T_{i-1}},
\end{equation*}
as less than $N$ balls need to be added to the tree in total. Note that we
have not made sure yet that the bins are not contacted by other balls; ${\cal
E}_3$ is concerned with constructing the tree as a subgraph of $G_{\A}(t)$ only.

For ${\cal E}_3$ to happen, we also need that all balls that are added to the
tree contact previously isolated bins. Hence, in total fewer than $N$ u.i.r.\
choices need to hit different bins from a subset of size $n e^{-(1+o(1))C
n/T_{i-1}}$. This probability can be bounded by
\begin{equation*}
\left(\frac{n e^{-(1+o(1))C n/T_{i-1}}-N}{n}\right)^N\sr{eq:N_weak}{\in}
e^{-(1+o(1))CN n/T_{i-1}}.
\end{equation*}

Now, after constructing the tree, we need to make sure that it is indeed
the induced subgraph of ${\cal N}_R^{(2D)}$ in $G_{\A}(i)$, i.e., no further
edges connect to any nodes in the tree. Denote this event by ${\cal E}_4$. As
we already ``used'' all edges of balls inside the tree and there are no more
than $C n^2/T_{i-1}$ edges created by balls outside the tree, ${\cal E}_4$
happens with probability at least
\begin{equation*}
\left(1-\frac{N}{n}\right)^{C n^2/T_{i-1}}\in e^{-(1+o(1))CN n/T_{i-1}}.
\end{equation*}

Combining all factors, we obtain that
\begin{eqnarray*}
p&\geq & P[{\cal E}_1]\cdot P[{\cal E}_2\,|\,{\cal E}_1]\cdot 
P[{\cal E}_3\,|\,{\cal E}_1\wedge {\cal E}_2]\cdot
P[{\cal E}_4\,|\,{\cal E}_1\wedge {\cal E}_2 \wedge {\cal E}_3]\\
&\in &\left(1-\frac{1}{n^c}\right)^2 e^{-(1+o(1))(C+1)N n/T_{i-1}}
e^{-(1+o(1))CN n/T_{i-1}}\\
&= & e^{-(1+o(1))(2C+1)N n/T_{i-1}}\\
&\sr{eq:N_strong}{\subset} & 2N
e^{-(1+o(1))(2C+1)(2\Dbi_i\Dba_i)^{2^t}n/T_{i-1}}e^{(1+o(1))C n/T_{i-1}}\\
&\subseteq & 2N
e^{-(1+o(1))(2C+1)\big(4L(2\Dba_1n/T_{i-1})^2\big)^{2^t}n/T_{i-1}}
e^{(1+o(1))C n/T_{i-1}}\\
&\subseteq & 2N2^{-\big(n/T_0 (n/T_{i-1})^3\big)^{2^t}}e^{(1+o(1))C
n/T_{i-1}}\\
&\subseteq & \frac{2N T_i}{n} e^{(1+o(1))C n/T_{i-1}}.
\end{eqnarray*}

We conclude that the expected value of the random variable $X$ counting the
number of disjoint $((\Dbi_1,\ldots,\Dbi_i),(\Dba_1,\ldots,\Dbi_i),2^t)$-trees is
lower bounded by $E[X]> 2T_i$, as at least $e^{-(1+o(1))C n/T_{i-1}}n$ isolated
bins are left that may serve as root of (not necessarily disjoint) trees and each
tree contains less than $N$ bins.

Finally, having fixed $G_{\A}(i-1)$, $X$ becomes a function of w.h.p.\ at most
$\BO(n^2/T_{i-1})\subseteq\BO(n^2/T_{t-1})\subseteq
\BO(n\log(n/T_t))\subseteq\BO(n\log n)$ u.i.r.\ chosen bins contacted by the
balls in round $i$. Each of the corresponding random variables may change the
value of $X$ by at most three: An edge insertion may add one tree or
remove two, while deleting an edge removes at most one tree and creates at most
two. Due to the prerequisite that $T_i\geq T_t\in \omega(\sqrt{n}\log n)$, we
have $E[X]\in \omega(\sqrt{n}\log n)$. Hence we can apply
Theorem~\ref{theorem:azuma} in order to obtain
\begin{equation*}
P\left[X<\frac{E[X]}{2}\right]\in e^{-\Omega\left(E[X]^2/(n\log
n)\right)}\subseteq n^{-\omega(1)},
\end{equation*}
proving the statement of the lemma.}

We see that the probability that layered trees occur falls exponentially
in their size to the power of $4\cdot 2^t$. Since $t$ is very small, i.e.,
smaller than $\log^* n$, this rate of growth is comparable to exponentiation by
a polynomial in the size of the tree. Therefore, one may expect that the
requirement of $T_t\in \omega(\sqrt{n}/\log n)$ can be maintained for values of
$t$ in $\Omega(\log^* n)$. Calculations reveal that even $t\in (1-o(1))\log^*
n$ is feasible.

\begin{lemma}\label{lemma:calc}
Using the notation of Lemma~\ref{lemma:sufficient_trees}, it holds for
\begin{equation*}
t\leq t_0(n,L)\in (1-o(1))\log^* n-\log^*L
\end{equation*}
that $T_t\in \omega(\sqrt{n}/\log n)$.
\end{lemma}
\proof{Denote by
\begin{equation*}
^ka:=\left.a^{a^{\cdot^{\cdot^{\cdot^{a}}}}}\right\}k\in \N\mbox{ times}
\end{equation*}
the tetration with basis $a:=2^{4\cdot 2^t}$, and by $\log^*_a x$ the smallest
number such that $^{(\log^*_a x)}a\geq x$. By definition, we have that
\begin{eqnarray*}
\frac{n}{T_t}&=&2^{(n/T_{t-1})^{4\cdot 2^t}}\\
&=&2^{\big(2^{(n/T_{t-2})^{4\cdot 2^t}}\big)^{4\cdot 2^t}}\\
&=&2^{2^{4\cdot 2^t\cdot(n/T_{t-2})^{4\cdot 2^t}}}\\
&=&2^{a^{(n/T_{t-2})^{4\cdot 2^t}}}\\
&=&2^{a^{a^{(n/T_{t-3})^{4\cdot 2^t}}}}.
\end{eqnarray*}
Repeating this computation inductively, we obtain
\begin{equation*}
\log\left(\frac{n}{T_t}\right)\leq
{^{(t+\log^*_a (T_0/n))}a}.
\end{equation*}
Applying $\log^*$, we get the sufficient condition
\begin{equation}\label{eq:condition}
\log^*\left({^{(t+\log^*_a (T_0/n))}a}\right)\leq \log^*n -3,
\end{equation}
since then
\begin{eqnarray*}
\frac{n}{T_t}\leq {^{\log^*(n/T_t)}2}\leq {^{(\log^*n-2)}}2\leq
\log n\in o\left(\frac{\sqrt{n}}{\log n}\right).
\end{eqnarray*}

Assume that $a,k\geq 2$. We estimate
\begin{eqnarray*}
\log^*({^ka} (1+\log a))&=&1+\log^*(\log({^ka} (1+\log a)))\\
&=&1+\log^*({^{(k-1)}a}\log a +\log(1+\log a))\\
&\leq &1+\log^*({^{(k-1)}a}(1+\log a)).
\end{eqnarray*}
By induction on $k$, it follows that
\begin{equation*}
\log^*({^ka})\leq k-1+\log^*(a(1+\log a))\leq k+\log^* a.
\end{equation*}
This implies
\begin{eqnarray*}
\log^*\left({^{(t+\log^*_a (T_0/n))}a}\right)
&\leq & t+\log^*_a (T_0/n)+\log^* a\\
&\leq & t+\log^*L+\BO(1)+\log^*a\\
&\subset & (1+o(1))t+\log^* L+\BO(1).
\end{eqnarray*}
Thus, slightly abusing notation, Inequality~\eqref{eq:condition} becomes
\begin{equation*}
(1+o(1))t+\log^* L+\BO(1)\leq \log^* n-\BO(1),
\end{equation*}
which is equivalent to the statement of the lemma.}

In light of Corollary~\ref{coro:algo_constant}, this interplay between $L$ and
$t$ is by no means arbitrary. If for any $r\in \N$ one accepts a
maximum bin load of $\log^{(r)}n/\log^{(r+1)} n + r +\BO(1)$, where
$\log^{(r)}$ denotes the $r$ times iterated logarithm, Problem~\ref{prob:bib}
can be solved in $r+\BO(1)$ rounds by $\A_b(r)$.

Since now we know that critical subgraphs occur frequently for specific
algorithms, next we prove that this subclass of algorithms is as powerful as
acquaintance algorithms of certain bounds on time and message complexity.

\begin{lemma}\label{lemma:simplify_algo}
Suppose the acquaintance Algorithm $\A$ solves Problem \ref{prob:bib} within
$t\leq t_0(n,L)$, $L\in \N$, rounds w.h.p.\ ($t_0$ as in
Lemma~\ref{lemma:calc}), sending w.h.p.\ at most $\BO(n)$ messages in total and
$\polylog n$ messages per node. Then, for sufficiently large $n$, a constant
$\Dba_1$ and an oblivious-choice algorithm $\A'$ with regard to the set of
parameters specified in Lemma~\ref{lemma:sufficient_trees} exists that sends
at most $\BO(n^2/T_{i-1})$ messages in round $i\in \{1,\ldots,t\}$ w.h.p.,
terminates at the end of round $t$, and is w.h.p.\ equivalent to $\A$.
\end{lemma}
\proof{Observe that $\A$ has only two means of disseminating information: Either,
balls can randomly connect to unknown bins, or they can send information to bins
known from previous messages. Thus, any two nodes at distance larger than $2^t$
from each other in $G_{\A}(i-1)$ must act independently in round $i$. Since
degrees are at most $\polylog n$  w.h.p., w.h.p.\ no ball knows more than
$(\polylog n)^{2^t}\subseteq n^{o(1)}$ bins (recall that $t_0(n,l)\leq \log^* n
-2$). Assume that in a given round a ball chooses $k$ bins, excluding the ones
of which he already obtained the global address. If it contacted $k':=\lceil
3ck\rceil$ bins u.i.r.\ and dropped any drawn bin that it already knows
(including repetitions in the current round), it would make with probability at
most
\begin{equation*}
\sum_{j=k'-k+1}^{k'}
\binom{k'}{j}\left(1-\frac{1}{n^{1-o(1)}}\right)^{k'-j}\frac{1}{n^{(1-o(1))j}}
\subset n^{-2(1-o(1))c}\polylog n\subset n^{-c}
\end{equation*}
less than $k$ new contacts. Thus, we may modify $\A$ such that it chooses
$\BO(k)$ bins u.i.r.\ whenever it would contact $k$ bins randomly. This can be
seen as augmenting $G_{\A}(i)$ by additional edges. By ignoring these edges, the
resulting algorithm $\A'$ is capable of (locally) basing its decisions on the
probability distribution of graphs $G_{\A}(i)$. Hence, Condition~$(ii)$ from
Definition~\ref{def:oblivious_choice} is met by the modified algorithm.

Condition~$(i)$ forces $\A'$ to terminate in round $t$ even if $\A$ does not.
However, since $\A$ must terminate in round $t$ w.h.p., balls may choose
arbitrarily in this case, w.h.p.\ not changing the output compared to $\A$. On
the other hand, we can certainly delay the termination of $\A$ until round $t$
if $\A$ would terminate earlier, without changing the results. Thus, it remains
to show that we can further change the execution of $\A$ during the first $t$
rounds in a way ensuring Condition~$(iii)$ of the definition, while maintaining
the required bound on the number of messages.

To this end, we modify $\A$ inductively, where again in each round for some balls
we increase the number of randomly contacted bins compared to an execution of
$\A$; certainly this will not affect Conditions~$(i)$ and $(ii)$ from
Definition~\ref{def:oblivious_choice}, and $\A'$ will exhibit the same output
distribution as $\A$ (up to a fraction of $1/n^c$ of the executions) if $\A'$
also ignores additional edges when placing the balls at the end of round $t$.

Now, assume that the claim holds until round $i-1\in \{0,\ldots,t-1\}$. In round
$i\in \{1,\ldots,t\}$, balls in depth at most $2^t$ of disjoint
$((\Dbi_1,\ldots,\Dbi_{i-1}),(\Dba_1,\ldots,\Dba_{i-1}),2^t)$-trees in
$G_{\A'}(i)$ are in distance at least $2^{t+1}$ from each other, i.e., they
must decide mutually independently on the number of bins to contact.
Consequently, since $\A$ has less information at hand than $\A'$, these balls
would also decide independently in the corresponding execution of $\A$. For
sufficiently large $n$, (the already modified variant of) $\A$ will w.h.p.\ send
at most $C n$ messages. Hence, the balls that are up to depth $2^t$ of such a
tree send together in expectation less than $2n C/T_{i-1}$ messages, since
otherwise Theorem~\ref{theorem:chernoff} and the fact that $T_{i-1}\in
\omega(\log n)$ would imply that w.h.p.\ at least $(2-o(1))C n$ messages would be
sent by $\A$ in total. Consequently, by Markov's Inequality, with independent
probability at least $1/2$, all balls in depth $2^t$ or less of a layered
$((\Dbi_1,\ldots,\Dbi_{i-1}),(\Dba_1,\ldots,\Dba_{i-1}),2^t)$-tree in union send
no more than $4C n/T_{i-1}$ messages in round $i$. Using
Theorem~\ref{theorem:chernoff} again, we conclude that for w.h.p.\
$\Omega(T_{i-1})$ many trees it holds that none of the balls in depth at most
$2^t$ will send more than $4C n/T_{i-1}$ messages to randomly chosen bins.

When executing $\A'$, we demand that each such ball randomly contacts
\emph{exactly} that many bins, i.e., with $\Dba_1:=4C$ Condition $(iii)$ of
Definition~\ref{def:oblivious_choice} is met. By Theorem~\ref{theorem:chernoff},
this way w.h.p.\ at most $\BO(n+n^2/T_{i-1})=\BO(n^2/T_{i-1})$ messages are sent
in round $i$ as claimed. Moreover, the new algorithm can ensure to follow the
same probability distribution of bin contacts as $\A$, simply by ignoring the
additional random choices made. This completes the induction step and thus the
proof.}

The final ingredient is to show that randomization is insufficient to deal with
the highly symmetric topologies of layered trees. In particular, balls that
decide on bins despite not being aware of leaves cannot avoid risking to choose
the root bin of the tree. If all balls in a tree where bin degrees are large
compared to ball degrees decide, this results in a large load of the root bin.

\begin{figure}[t!]
\begin{center}
\includegraphics[width=\textwidth]{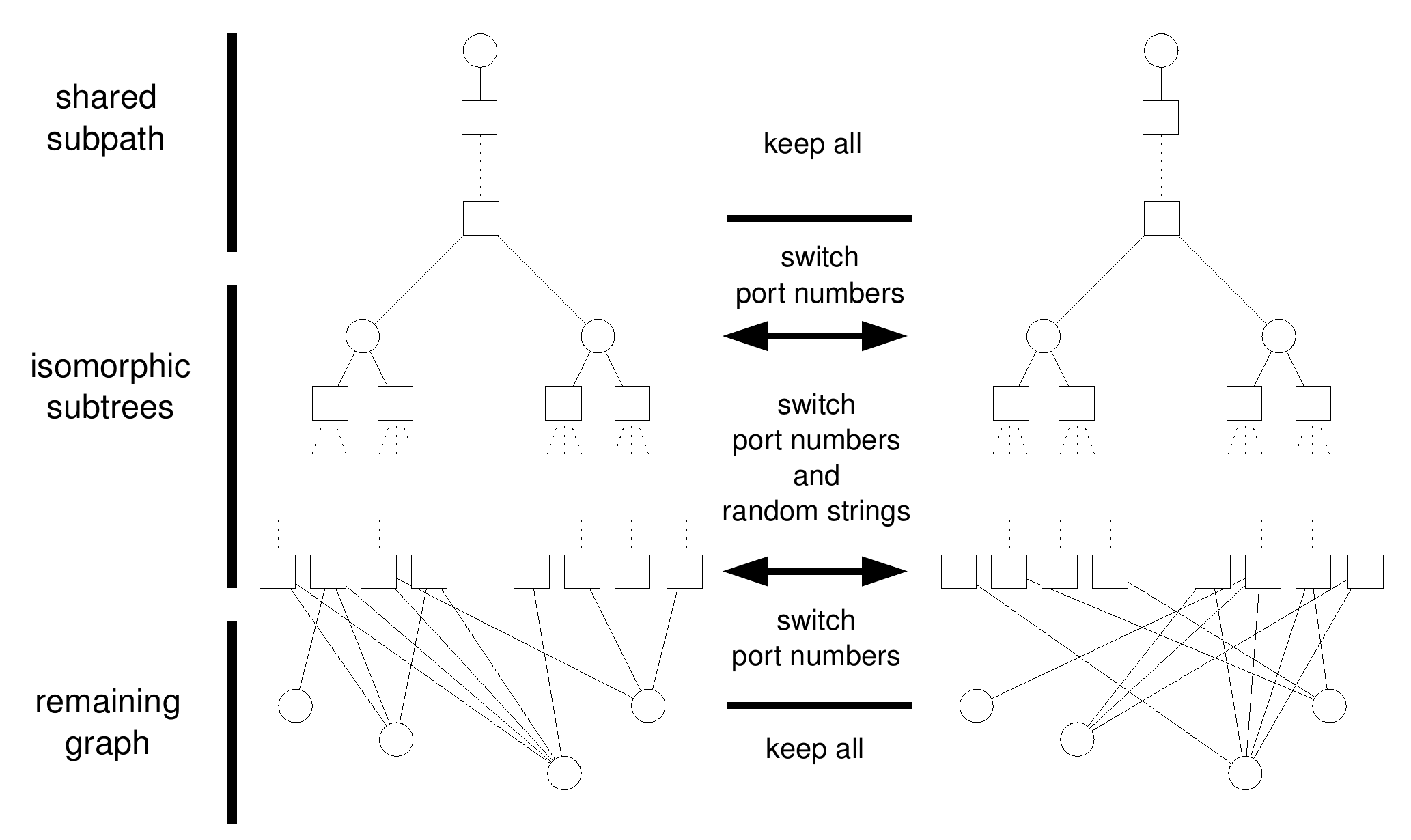}\label{fig:switch_tree}
\caption{Example for the effect of $h$ on the topology of $G_{\A}(t)$.
Switching port numberings and random labels of isomorphic subtrees and the port
numberings of their neighborhood, nodes in these subtrees essentially ``switch
their identity'' with their counterparts. Since the local view of the topmost
ball is completely contained within the tree, it cannot distinguish between the
two configurations.}
\end{center}
\end{figure}

\begin{lemma}\label{lemma:symmetric_choices}
Suppose after $t$ rounds of some Algorithm $\A$ ball $b$ is in depth at most
$2^t$ of a layered $(\Dbi,\Dba,2^t)$-tree of $t$ levels in $G_{\A}(t)$. We fix
the topology of the layered tree. Let $v$ be a bin in distance $d\leq 2^t$ from
$b$ in $G_{\A}(t)$ and assume that the edge sequence of the (unique) shortest
path from $b$ to $v$ is $e_1,\ldots,e_d$. If $b$ decides on a bin in round $t$,
the probability that $b$ places itself in $v$ depends on the sequence of rounds
$\ell_1,\ldots,\ell_d\in \{1,\ldots,t\}^d$ in which the edges $e_1,\ldots,e_d$
have been created only.
\end{lemma}
\proof{Observe that since $b$ is a ball, it must have an odd distance from the
root of the layered $(\Dbi,\Dba,2^t)$-tree it participates in. Thus, the
$2^t$-neighborhood of $b$ is a subset of the $(2^{t+1}-1)$-neighborhood of the
root of the layered tree. Therefore, this neighborhood is a balanced tree of
uniform ball degrees. Moreover, for all $i\in \{1,\ldots,t\}$, the number of
edges from $E_{\A}(i)$ balls up to distance $2^t$ from $b$ are adjacent to is the
same (including the balls that are leaves). Bin degrees only depend on the round
$i$ in which they have been contacted first and all their edges were created in
that round (cf.~Figure~\ref{fig:build_tree}).

Let $b_1,\ldots,b_n$ and $v_1,\ldots,v_n$ be global, fixed enumerations of the
balls and bins, respectively. Fix a topology $T$ of the $2^t$-neighborhood of $b$
with regard to these enumerations. Assume that $v$ and $w$ are two bins in $T$
for which the edges on the shortest paths from $b$ to $v$ resp.\ $w$ were added
in the rounds $\ell_1,\ldots,\ell_d$, $d\in \{1,\ldots,2^t\}$. Assume that $x$
and $y$ are the first distinct nodes on the shortest paths from $b$ to $v$ and
$w$, respectively. The above observations show that the subtrees of $T$ rooted at
$x$ and $y$ are isomorphic (cf.~Figure~\ref{fig:switch_tree}). Thus, a graph
isomorphism $f$ exists that ``exchanges'' the two subtrees (preserving their
orientation), fixes all other nodes, and fulfills that $f(v)=w$ and $f^2$ is the
identity. We choose such an $f$ and fix it. Denote by $p(b_i,v_j)\in
\{1,\ldots,n\}$, $i,j\in \{1,\ldots,n\}$, the port number $v_j$ has in the port
numbering of $b_i$ and by $p(v_i,b_j)$ the number $b_j$ has in the port numbering
of bin $v_i$. Similarly, $r(b_i)$ and $r(v_i)$, $i\in \{1,\ldots,n\}$, denote the
random inputs of $b_i$ and $v_i$, respectively. Using $f$, we define the
automorphism $h:S\to S$ on the set of tuples of possible port numberings and
random strings $(p(\cdot,\cdot),r(\cdot))$ by
\begin{equation*}
h((p(\cdot,\cdot),r(\cdot))):=(p(f(\cdot),f(\cdot)),r(f(\cdot))).
\end{equation*}

Set
\begin{equation*}
S_v:=\{(p(\cdot,\cdot),r(\cdot))\,|\,T\mbox{ occurs}\wedge b\mbox{ chooses }v\}
\subset S
\end{equation*}
and $S_w$ analogously. We claim that $h(S_v)=S_w$ (and therefore also
$h(S_w)=h^2(S_v)=S_v$). This means when applying $h$ to an element of $S_v$, the
topology $T$ is preserved and $b$ chooses $w$ instead of $v$. To see this,
observe that $\A$ can be interpreted as deterministic algorithm on the (randomly)
labeled graph where nodes $u$ are labeled $r(u)$ and edges $(u,u')$ are labeled
$(p(u,u'),p(u',u))$. Hence, $h$ simply switches the local views of $u$ and
$f(u)$ in that graph and a node $u$ takes the role of $f(u)$ and vice versa
(cf.~Figure~\ref{fig:switch_tree}). Thus, $b$ will choose $f(v)=w$ in the
execution with the new labeling. On the other hand, $T$ is preserved because we
chose the function $f$ in a way ensuring that we mapped the two subtrees in $T$
that are rooted at $x$ and $y$ to each other by a graph isomorphism, i.e., the
topology with regard to the fixed enumerations $b_1,\ldots,b_n$ and
$v_1,\ldots,v_n$ did not change.

In summary, for any topology $T$ of the $2^t$-neighborhood of $b$ in a
layered $(\Dbi,\Dba,2^t)$-tree such that $b$ is connected to $v$ and $w$ by
shortest paths for which the sequences of round numbers when the edges were
created are the same, we have that $S_v=h(S_w)$. Since both port
numberings and random inputs are chosen independently, we conclude that
$P[(p(\cdot,\cdot),r(\cdot))\in S_v]=P[(p(\cdot,\cdot),r(\cdot))\in S_w]$, i.e.,
$b$ must choose $v$ and $w$ with equal probability as claimed.}

We are now in the position to prove our lower bound on the trade-off between
maximum bin load and running time of acquaintance algorithms.
\begin{theorem}\label{theorem:bib_lower}
Any acquaintance algorithm sending w.h.p.\ in total $\BO(n)$ messages and
at most $\polylog n$ messages per node either incurs w.h.p.\ a maximum bin load
of more than $L\in \N$ or runs for $(1-o(1))\log^* n-\log^* L$ rounds,
irrespective of the size of messages.
\end{theorem}
\proof{Assume that Algorithm $\A$ solves Problem~\ref{prob:acbib}
within at most $t\leq t_0\in (1-o(1))\log^* n-\log^* L$ rounds w.h.p.\ ($t_0$ as
in Lemma~\ref{lemma:calc}). Thus, due to Lemma~\ref{lemma:simplify_algo} a
constant $\Dba_1$ and an oblivious-choice Algorithm $\A'$ with regard to the
parameters from Lemma~\ref{lemma:sufficient_trees} whose maximum bin load
is w.h.p.\ the same as the one of $\A$ exist. In the following, we use the
notation from Lemma~\ref{lemma:sufficient_trees}.

Suppose that $R$ is the root bin of a layered $(\Dbi,\Dba,2^t)$-tree in
$G_{\A'}(t)$. According to Lemma~\ref{lemma:symmetric_choices}, for all balls $b$
in distance up to $2^t$ from $R$, the probability $p$ to choose a bin $v$ solely
depends on the sequence $s(b,v)=(s_1,\ldots,s_d)$ of round numbers when the edges
on the shortest path from $R$ to $b$ were created. Set
$S:=\bigcup_{i=1}^{2^{t-1}}S_{2i-1}$, where $S_d$ denotes the set of round
sequences $s=(s_1,\ldots,s_d)$ of (odd) length $d(s):=d$ from balls to bins
(inside the tree). Denote for $s\in S$ by $p(s)$ the probability that a ball $b$
within distance $2^t$ from $R$ decides on (any) bin $v$ with $s(b,v)=s$ and by
$X$ the random variable counting the number of balls deciding on $R$. Recall
that for any $i\in \{1,\ldots,t\}$, we have $\Dbi_i=2L\Dba_i$. We compute
\begin{eqnarray*}
E(X)&=& \sum_{s\in S}p(s)
\frac{|\{b\in V_{\circ}\,|\,s(b,R)=s\}|}{|\{v\in V_{\sqcup}\,|\,s(b,v)=s\}|}\\
&=& \sum_{s\in
S}p(s)\frac{\Dbi_{s_1}\Pi_{i=1}^{\lfloor d(s)/2\rfloor}
\Dba_{s_{2i}}\Dbi_{s_{2i+1}}}{\Dba_{s_1}\Pi_{i=1}^{\lfloor d(s)/2\rfloor}
\Dbi_{s_{2i}}\Dba_{s_{2i+1}}}\\
&=& \sum_{s\in S}p(s)2L\\
&=& 2L,
\end{eqnarray*}
as each ball must decide with probability $1$ on a bin within distance $2^t$
(Condition $(ii)$ of Definition~\ref{def:oblivious_choice}).

On the other hand, the maximum possible load of $R$ is the number of balls
up to depth $2^t$ of the tree, which we observed to be less than
$(2\Dbi_t\Dba_t)^{2^t}\in \BO((n/T_{t-1})^2)\subset (\log n)^2$. We infer that
for sufficiently large $n$ we have that $P[X>L]> 1/(\log n)^2$, since otherwise
\begin{equation*}
2L=E(X)\leq (1-P[X> L])L+P[X>L](\log n)^2<2L.
\end{equation*}

As Lemma~\ref{lemma:calc} states that the number of disjoint layered
$(\Dbi,\Dba,2^t)$-trees is w.h.p.\ at least $T_t\in \omega(\sqrt{n}/\log n)$, we
have for the random variable $Y$ counting the number of roots of such trees that
get a bin load of more than $L$ that
\begin{equation*}
E(Y)\geq \left(1-\frac{1}{n^c}\right)P[X>L]T_t \in \omega(\log n).
\end{equation*}
Recall that Lemma~\ref{lemma:symmetric_choices} holds for fixed topologies of
the trees, i.e., the estimates for $P[X>L]$ and thus $E(Y)$ follow after fixing
the topology up to distance $2^{t+1}$ from all the roots first. Thus, whether a
root bin gets a load of more than $L$ is independent of the other roots' loads,
since there is no communication between the involved balls. We conclude that
we can apply Theorem~\ref{theorem:chernoff} to $Y$ in order to see that w.h.p.\
$Y>0$, i.e., w.h.p.\ $\A'$ incurs a maximum bin load larger than $L$. Because
$\A$ and $\A'$ are w.h.p.\ equivalent, the same must be true for $\A$, proving
the claim.}

\subsection{Generalizations}
For ease of presentation, the proof of the lower bound assumed that bins do not
contact other bins. This is however not necessary.
\begin{corollary}\label{coro:bib_lower_bins}
Theorem \ref{theorem:bib_lower} holds also if bins may directly exchange
messages freely.
\end{corollary}
\textbf{Proof Sketch.} 
The presented technique is sufficient for the more
general case, as can be seen by the following reasoning. To adapt the proof, we
have to consider trees similar to layered $(\Dba,\Dbi,2^t)$-trees, where now also
bins form edges. Therefore, also bins may create an in each round exponentially
growing number of edges to other bins. However, the probability that a bin is
the root of such a tree structure in round $i$ will still be lower bounded by
$2^{-(n/T_{i-1})^{f(t)}}$, where $f(t)$ is some function such that
$\log^*(f(t))\in \log^* t +\BO(1)$ and $T_{i-1}$ is a lower bound on the number
of such roots in round $i-1$. Hence, Lemmas~\ref{lemma:sufficient_trees}
and~\ref{lemma:calc} can be changed accordingly. From that point on, the
remainder of the proof can be carried out analogously.\qed

It is important to be aware that this holds only as long as bins initially
identify each other according to u.i.r.\ port numberings as well. If bins are
aware of a globally consistent labeling of all bins, an asymmetric algorithm
can be executed, as bins may support balls in doing asymmetric random choices.

Similarly, the upper bound on the number of messages individual nodes send can
be relaxed.
\begin{corollary}\label{coro:bib_lower_mess}
Theorem \ref{theorem:bib_lower} holds also if nodes send at most $\lambda n$
messages in total, where $\lambda \in [0,1)$ is a constant.
\end{corollary}
\textbf{Proof Sketch.} The critical point of the argumentation is in the proof
of Lemma~\ref{lemma:simplify_algo}, where we replace nodes' new contacts by
u.i.r.\ chosen bins. For large numbers of messages, we can no longer guarantee
w.h.p.\ that increasing the number of balls' random choices by a constant factor
can compensate for the fact that balls will always contact different bins with
each additional message. Rather, we have to distinguish between nodes sending
many messages and ones sending only few (e.g.\ $\polylog n$). Only to the
latter we apply the replacement scheme.

Of course, this introduces new difficulties. For instance, we need to observe
that still a constant fraction of the bins remains untouched by balls sending
many messages for the proof of Lemma~\ref{lemma:sufficient_trees} to hold. The
worst case here would be that $\BO(1)$ nodes send $\lambda n$ messages during the
course of the algorithm, since the bound of $\BO(n)$ total messages w.h.p.\ must
not be violated. Thus, the probability that a bin is not contacted by such a ball
is lower bounded by
\begin{equation*}
1-(1-\lambda)^{\BO(1)}\subseteq \Omega(1).
\end{equation*}
Using standard techniques, this gives that still many bins are never contacted
during the course of the algorithm w.h.p. Similarly, we need to be sure that
sufficiently many of the already constructed trees are not contacted by
``outsiders''; here we get probabilities that fall exponentially in the size of
such a tree, which is sufficient for the applied techniques.

Another aspect is that now care has to be taken when applying
Theorem~\ref{theorem:azuma} to finish the proof of
Lemma~\ref{lemma:sufficient_trees}. The problem here is that the random variable
describing the edges formed by a ball of large degree is not the product of
independent random variables. On the other hand, treating it as a single variable
when applying the theorem, it might affect all of the layered trees, rendering
the bound from the theorem useless. Thus, we resort to first observing that not
too many nodes will send a lot of messages, then fixing their random choices,
subsequently bounding the expected number of layered trees conditional to these
choices already being made, and finally applying Theorem~\ref{theorem:azuma}
merely to the respective random variable depending on the edges created
u.i.r.\ by balls with small degrees only.\qed

Note that if we remove the upper bound on the number of messages a single node
might send entirely, there is a trivial solution:
\begin{enumerate}
  \item With probability, say, $1/\sqrt{n}$, a ball contacts $\sqrt{n}$ bins.
  \item These balls perform a leader election on the resulting graph (using
  random identifiers).
  \item Contacting all bins, the leader coordinates a perfect distribution of
  the balls.
\end{enumerate}
In the first step $\BO(n)$ messages are sent w.h.p. Moreover, the subgraph
induced by the created edges is connected and has constant diameter w.h.p.
Hence Step~2, which can be executed with $\BO(n)$ messages w.h.p., will result
in a single leader, implying that Step~3 requires $\BO(n)$ messages as well.
However, this algorithm introduces a central coordination instance. If this was
a feasible solution, there would be no need for a parallel balls-into-bins
algorithm in the first place.

In light of these results, our lower bound results essentially boil down to the
following. Any acquaintance algorithm that guarantees w.h.p.\ both small bin
loads and asymptotically optimal $\BO(n)$ messages requires $(1-o(1))\log^* n$
rounds.

\section{Constant-Time Solutions}\label{sec:upper}

Considering Theorem~\ref{theorem:bib_lower},
Corollaries~\ref{coro:bib_lower_bins} and~\ref{coro:bib_lower_mess}, and
Theorem~\ref{coro:constant}, two questions come to mind.

\begin{itemize}
  \item Does the lower bound still hold if random choices may be
  \emph{asymmetric}, i.e., non-uniform choice distributions are possible?
  \item What happens if the bound of $\BO(n)$ on the total number of messages
  is relaxed?
\end{itemize}
In this section, we will discuss these issues.

\subsection{An Asymmetric Algorithm}
In order to answer the first question, we need to specify precisely what
dropping the assumption of symmetry means.
\begin{problem}[Asymmetric Balls-into-Bins]\label{prob:asbib}
An instance of Problem~\ref{prob:bib} is an \emph{asymmetric parallel
balls-into-bins problem}, if balls identify bins by globally unique addresses
$1,\ldots,n$. We call an algorithm solving this problem \emph{asymmetric}.
\end{problem}
``Asymmetric'' here means that biased random choices are permitted. This is
impossible for symmetric or acquaintance algorithms, where the uniformly random
port numberings even out any non-uniformity in the probability distribution of
contacted port numbers.

In this subsection, we will show that asymmetric algorithms can indeed obtain
constant bin loads in constant time, at asymptotically optimal communication
costs. Note that for asymmetric algorithms, we can w.l.o.g.\ assume that $n$ is a
multiple of some number $l\in o(n)$, since we may simply opt for ignoring
negligible $n-l\lfloor n/l\rfloor$ bins. We will use this observation in the
following. We start by presenting a simple algorithm demonstrating the basic idea
of our solution. Given $l\in \BO(\log n)$ that is a factor of $n$, $\A_1(l)$ is
defined as follows.
\begin{enumerate}
  \item Each ball contacts one bin chosen uniformly at random from the set
  $\{i l\,|\,i\in \{1,\ldots,n/l\}\}$.
  \item Bin $i l$, $i\in\{1,\ldots,n/l\}$, assigns up to $3l$ balls to the bins
  $(i-1)l+1,\ldots,il$, such that each bin gets at most three balls.
  \item The remaining balls (and the bins) proceed as if executing the symmetric
  Algorithm~$\A_b^2$, however, with $k$ initialized to
  $k(1):=2^{\alpha l}$ for an appropriately chosen constant $\alpha>0$.
\end{enumerate}
Essentially, we create buckets of non-constant size $l$ in order to ensure
that the load of these buckets is slightly better balanced than it would be the
case for individual bins. This enables the algorithm to place more than a
constant fraction of the balls immediately. Small values of $l$ suffice for this
algorithm to terminate quickly.
\begin{lemma}\label{lemma:algo_simple}
Algorithm~$\A_1(l)$ solves Problem~\ref{prob:asbib} with a maximum bin load of
three. It terminates within $\log^* n - \log^* l + \BO(1)$ rounds w.h.p.
\end{lemma}
\proof{Let for $i\in \N_0$ $Y^i$ denote the random variables counting the
number of bins receiving at least $i$ messages in Step~1. From Lemma
\ref{lemma:balls_bins_negative} we know that Theorem~\ref{theorem:chernoff}
applies to these variables, i.e., $|Y^i-E(Y^i)|\in \BO\left(\log n +
\sqrt{E(Y^i)\log n}\right)$ w.h.p. Consequently, we have that
the number $Y^i-Y^{i+1}$ of bins receiving exactly $i$ messages differs by at
most $\BO\left(\log n + \sqrt{\max\{E(Y^i),E(Y^{i+1})\}\log n}\right)$ from its
expectation w.h.p. Moreover, Theorem~\ref{theorem:chernoff} states that these
bins receive at most $l+\BO(\sqrt{l\log n}+\log n)\subset \BO(\log n)$
messages w.h.p., i.e., we need to consider only values of $i\in \BO(\log n)$.

Thus, the number of balls that are not accepted in the first phase is bounded by
\begin{eqnarray*}
&& \sum_{i=3l+1}^n (i-3l)\left(Y^i-Y^{i+1}\right)\\
&\in & \sum_{i=3l+1}^{\BO(\log n)} (i-3l)\,E\!\left(Y^i-Y^{i+1}\right)
+\BO\left(\sqrt{n \log n}\right)\\
&\subseteq &\frac{n}{l}\sum_{i=3l+1}^{\BO(\log n)}(i-3l)\binom{n}{i}
\left(\frac{l}{n}\right)^i\left(1-\frac{l}{n}\right)^{n-i}
+\BO\left(\sqrt{n \log n}\right)\\
&\subseteq &
\frac{n}{l}\sum_{i=3l+1}^{\BO(\log n)}(i-3l)
\left(\frac{e l}{i}\right)^i +\BO\left(\sqrt{n \log n}\right)\\
&\subseteq &
\frac{n}{l}\sum_{j=1}^{\infty} j l
\left(\frac{e}{3}\right)^{(j+2)l}+\BO\left(\sqrt{n \log n}\right)\\
&\subseteq &
\BO\left(\left(\frac{e}{3}\right)^{2l}n+\sqrt{n \log n}\right)\\
&\subseteq &\left(\frac{3}{e}\right)^{-(2-o(1))l}n+\BO\left(\sqrt{n \log
n}\right)
\end{eqnarray*}
w.h.p., where in the third step we used the inequality $\binom{n}{i}\leq
(en/i)^i$.

Thus, w.h.p.\ at most $2^{-\Omega(l)}n+\BO\left(\sqrt{n \log n}\right)$ balls are
not assigned in the first two steps. Hence, according to
Corollary~\ref{coro:quick}, we can deal with the remaining balls within $\log^* n
-\log^* l+\BO(1)$ rounds by running $\A_b^2$ with $k$ initialized to $2^{\alpha
l}$ for $\alpha \in (2-o(1))\log (3/e)$ when executing $\A_s$. We conclude that
$\A_1(l)$ will terminate after $\log^* n - \log^* l + \BO(1)$ rounds w.h.p.\ as
claimed.}

In particular, if we set $l:=\log^{(r)} n$, for any $r\in \N$, the algorithm
terminates within $r+\BO(1)$ rounds w.h.p. What is more, this algorithm
provides the means for an optimal solution of Problem~\ref{prob:sidt}, as stated
in Theorem~\ref{theorem:optimal}.\vspace{0.3\baselineskip}
\noindent\textbf{Proof of Theorem~\ref{theorem:optimal}.} Algorithm $\A_1(l)$
offers a maximum bin load of three in constant time for e.g.\ $l:=\log n$. This
algorithm can be executed in parallel for each destination to distribute the
messages almost evenly among the nodes.\footnote{To avoid congestion at some
nodes, we order the nodes globally and send messages for the $k^{th}$ node to
the subset of nodes $\{(i l+k) \mbox{\;mod\,} n\,|\,i\in \{0,\ldots,n-1\}\}$ in
the first step of the algorithm.} Afterwards three rounds suffice to
deliver all messages to their destinations w.h.p.

The problem with this scheme is that $\A_1(l)$ assumes that balls choose
destinations independently, while we need to ensure that only constantly many
messages need to be sent over each link. This can be solved by two steps. First,
nodes send each held message to a uniformly chosen neighbor, such that each
neighbor receives exactly one message. This way, the number of messages that a
node $v\in V$ holds for any node $w\in V$ becomes a random variable $X_v^w$ that
is the sum of the variables indicating whether node $u\in V$ sends a message to
$v$ that is destined for $w$. These variables are independent, and by definition
of Problem~\ref{prob:sidt}, their expectations must sum up to exactly one. Hence,
Theorem~\ref{theorem:chernoff} states that, after one round, w.h.p.\ no node
will have more than $\BO(\log n)$ many messages destined to a single node.

Subsequently, nodes distribute messages (i.e., balls) that should contact u.i.r.\
nodes (i.e., bins) from a subset $S=\{(i l+k) \mbox{\;mod\,} n\,|\,i\in
\{0,\ldots,n-1\}\}$ of $n/l$ nodes according to the first step of $\A_1(l)$
uniformly among the nodes from $S$, but minimizing the maximum number of messages
sent along each link; they also avoid sending two messages with the same
destination to the same node, which is w.h.p.\ possible since $n/l\in \omega(\log
n)$. Applying Theorem~\ref{theorem:chernoff} once more, we see that this can be
done w.h.p.\ within a constant number of rounds, as nodes w.h.p.\ hold $\BO(n/l)$
messages destined to $S$. Even if we fix the $\BO(\log n)$ receivers of all other
messages destined to node $w\in V$ a node $v\in V$ holds, no node will receive a
single message with (conditional) probability larger than $1/(n-\BO(\log
n))\subset (1+o(1))/n$. Since this bound is independent from other messages'
(i.e., balls') destinations, the reasoning from Lemma~\ref{lemma:algo_simple}
applies for this slightly increased probabilities. Hence, the proof can be
carried out analogously, as the factor of $1+o(1)$ is asymptotically
negligible.\qed

However, the result of Lemma~\ref{lemma:algo_simple} is somewhat unsatisfactory
with respect to the balls-into-bins problem, since a subset of the bins has to
deal with an expected communication load of $l+\BO(1)\in \omega(1)$. Hence, we
want to modify the algorithm such that this expectation is constant.

To this end, assume that $l\in \BO(\log n/\log \log n)$ and $l^2$ is a factor of
$n$. Consider the following algorithm $\A_2(l)$ which assigns balls as
\emph{coordinators} of intervals of up to $l^2$ consecutive bins.
\begin{enumerate}
  \item With probability $1/l$, each ball picks one bin interval
  $I_j:=\{(j-1)l+1,\ldots,j l\}$, $j\in \{1,\ldots,n/l\}$, uniformly at random
  and contacts these bins. This message contains\linebreak $\lceil(c+2)\log
  n\rceil$ random bits.
  \item Each bin that receives one or more messages sends an acknowledgement to
  the ball whose random string represents the smallest number; if two or more
  strings are identical, no response is sent.
  \item Each ball $b$ that received acknowledgements from a contacted interval
  $I_j$ queries one u.i.r.\ chosen bin from each interval
  $I_{j+1},\ldots,I_{j+l-1}$ (taking indices modulo $n/l$) whether it has
  previously acknowledged a message from another ball; these bins respond
  accordingly. Ball $b$ becomes the coordinator of $I_j$ and all consecutive
  intervals $I_{j+1},\ldots,I_{j+k}$, $k<l$ such that none of these intervals
  has already responded to another ball in Step~2.
\end{enumerate}
The algorithm might miss some bin intervals, but overall most of the bins will
be covered.
\begin{lemma}\label{lemma:algo_bins}
When $\A_2(l)$ terminates after a constant number of rounds, w.h.p.\ all but
$2^{-\Omega(l)}n$ bins have a coordinator. The number of messages sent or
received by each ball is constant in expectation and at most $\BO(l)$. The number
of messages sent or received by each bin is constant in expectation and $\BO(\log
n/\log \log n)$ w.h.p. The total number of messages is $\BO(n)$ w.h.p.
\end{lemma}
\proof{The random strings chosen in Step~2 are unique w.h.p., since
the probability that two individual strings are identical is at most
$2^{-(c+2)\log n}=n^{-(c+2)}$ and we have $\binom{n}{2}<n^2$ different pairs of
balls. Hence, we may w.l.o.g.\ assume that no identical strings are received by
any bins in Step~2 of the algorithm.

In this case, if for some $j$ the bins in $I_j$ are not contacted in Steps~1
or~3, this means that $l$ consecutive intervals were not contacted by any
ball. The probability for this event is bounded by
\begin{equation*}
\left(1-\frac{1}{l}\cdot\frac{l}{n/l}\right)^{n}
=\left(1-\frac{l}{n}\right)^n< e^{-l},
\end{equation*}
Hence, in expectation less than $e^{-l}n/l$ intervals get no coordinator
assigned.

The variables indicating whether the $I_j$ have a coordinator are negatively
associated, as can be seen as follows. We interpret the first round as throwing
$n$ balls u.i.r.\ into $n$ bins, of which $n/l$ are labeled $I_1,\ldots,I_{n/l}$.
An interval $I_j$ has a coordinator exactly if one of the bins
$I_{j-l+1},\ldots,I_{j}$ (again, indices modulo $n/l$) receives a ball. We know
from Lemma~\ref{lemma:balls_bins_negative} that the indicator variables
$Y_i^1$ counting the non-empty bins are negatively associated; however, the third
step of its proof uses Statement~$(iii)$ from
Lemma~\ref{lemma:negative_association}, which applies to \emph{any} set of
increasing functions. Since maxima of increasing functions are increasing, also
the indicator variables
$\max\{Y_{I_{j-l+1}}^1,Y_{I_{j-l+2}}^1,\ldots,Y_{I_j}^1\}$ are negatively
associated.

Therefore, Theorem~\ref{theorem:chernoff} yields that the number of intervals
that have no coordinator is upper bounded by $\BO(e^{-l}n/l+\log n)$ w.h.p.
Consequently, w.h.p.\ all but $\BO(e^{-l}n+l\log n)\subseteq e^{-\Omega(l)}n$
bins are assigned a coordinator.

Regarding the communication complexity, observe that balls send at most
$\BO(l)$ messages and participate in the communication process with probability
$1/l$. In expectation, $n/l$ balls contact u.i.r.\ chosen bin intervals,
implying that bins receive in expectation one message in Step~1. Similarly, at
most $l$ balls pick u.i.r.\ bins from each $I_j$ to contact in Step~3. Since in
Step~3 at most $l-1$ messages can be received by bins, it only remains to
show that the bound of $\BO(\log n/\log \log n)$ on the number of
messages bins receive in Step~1 holds. This follows from the previous
observation that we can see Step~1 as throwing $n$ balls u.i.r.\ into $n$ bins,
where $n/l$ bins represent the $I_j$. For this setting the bound can be deduced
from Lemma~\ref{lemma:balls_bins_negative} and Theorem~\ref{theorem:chernoff}.
Finally, we apply Theorem~\ref{theorem:chernoff} to the number of balls
choosing to contact bins in Step~1 in order to see that $\BO(n)$ messages are
sent in total w.h.p.}

Finally, algorithm $\A(l)$ essentially plugs $\A_1(l)$ and $\A_2(l)$ together,
where $l\in \BO(\sqrt{\log n})$ and $l^2$ is a factor of $n$.
\begin{enumerate}
  \item Run Algorithm~$\A_2(l)$.
  \item Each ball contacts one bin, chosen uniformly.
  \item Each coordinator contacts the bins it has been assigned to by $\A_2(l)$.
  \item The bins respond with the number of balls they received a message from
  in Step~2.
  \item The coordinators assign (up to) three of these balls to each of
  their assigned bins. They inform each bin where the balls they received
  messages from in Step~2 need to be redirected.
  \item Each ball contacts the same bin as in Step~2. If the bin has a
  coordinator and the ball has been assigned to a bin, the bin responds
  accordingly.
  \item Any ball receiving a response informs the respective bin that it is
  placed into it and terminates.
  \item The remaining balls (and the bins) proceed as if executing Algorithm
  $\A_b^2$, however, with $k$ initialized to
  $k(1):=2^{\alpha l}$ for an appropriately chosen constant $\alpha>0$.
\end{enumerate}

\begin{theorem}
Algorithm $\A(l)$ solves Problem~\ref{prob:asbib} with a maximum bin load of
three. It terminates after $\log^*n -\log^* l +\BO(1)$ rounds w.h.p. Both balls
and bins send and receive a constant number of messages in expectation. Balls
send and receive at most $\BO(\log n)$ messages w.h.p., bins $\BO(\log n/\log
\log n)$ many w.h.p. The total number of messages is $\BO(n)$ w.h.p.
\end{theorem}
\proof{Lemma~\ref{lemma:algo_bins} states that all but $2^{-\Omega(l)}n$ bins
have a coordinator. Steps~2 to~7 of $\A(l)$ emulate Steps~1 and~2 of
Algorithm~$\A_1(l)$ for all balls that contact bins having a coordinator. By
Lemma~\ref{lemma:algo_simple}, w.h.p.\ all but $2^{-\Omega(l)}n$ of the balls
could be assigned if the algorithm would be run completely, i.e., with all bins
having a coordinator. Since w.h.p.\ only $2^{-\Omega(l)}n$ bins have no
coordinator and bins accept at most three balls, we conclude that w.h.p.\ after
Step~7 of $\A(l)$ merely $2^{-\Omega(l)}n$ balls have not been placed into bins.
Thus, analogously to Lemma~\ref{lemma:algo_simple}, Step~8 will require at most
$\log^* n- \log^* l +\BO(1)$ rounds w.h.p. Since Steps~1 to~7 require constant
time, the claimed bound on the running time follows.
The bounds on message complexity and maximum bin load are direct consequences of
Corollary~\ref{coro:quick}, Lemma~\ref{lemma:algo_bins}, the
definition of $\A(l)$, and the bound of $\BO(\sqrt{\log n})$ on $l$.}

Thus, choosing $l=\log^{(r)}n$ for any $r\in \N$, Problem~\ref{prob:asbib} can
be solved within $r+\BO(1)$ rounds.
\subsection{Symmetric Algorithm Using \texorpdfstring{$\omega(n)$}{omega(n)}
Messages}
A similar approach is feasible for symmetric algorithms if we permit $\omega(n)$
messages in total. Basically, Algorithm $\A(l)$ relied on asymmetry to assign
coordinators to a vast majority of the bins. Instead, we may settle for
coordinating a constant fraction of the bins; in turn, balls will need to send
$\omega(1)$ messages to find a coordinated bin with probability $1-o(1)$.

Consider the following Algorithm $\A_{\omega}(l)$, where $l\leq n/\log n$ is
integer.
\begin{enumerate}
  \item With probability $n/l$, a ball contacts a uniformly random subset of
  $l$ bins.
  \item Each bin receiving at least one message responds to one of these
  messages, choosing arbitrarily. The respective ball is the coordinator of the
  bin.
\end{enumerate}
This simple algorithm guarantees that a constant fraction of the bins will be
assigned to coordinators of $\Omega(l)$ bins.
\begin{lemma}\label{lemma:many_mess}
When executing $\A_{\omega}(l)$, bins receive at most $\BO(\log n /\log
\log n)$ messages w.h.p. In total $\BO(n)$ messages are sent w.h.p. W.h.p., a 
constant fraction of the bins is assigned to coordinators of $\Omega(l)$ bins.
\end{lemma}
\proof{Theorem~\ref{theorem:chernoff} states that in Step~1 w.h.p.\
$\Theta(n/l)$ balls decide to contact bins, i.e., $\Theta(n)$ messages are sent.
As before, the bound on the number of messages bins receive follows from
Lemma~\ref{lemma:balls_bins_negative} and Theorem~\ref{theorem:chernoff}. Using
Lemma~\ref{lemma:balls_bins_negative} and Theorem~\ref{theorem:chernoff} again,
we infer that w.h.p.\ a constant fraction of the bins receives at least one
message. Thus, $\Theta(n/l)$ balls coordinate $\Theta(n)$ bins, implying that
also $\Theta(n)$ bins must be coordinated by balls that are responsible for
$\Omega(l)$ bins.}

Permitting communication exceeding $n$ messages by more than a constant factor,
this result can be combined with the technique from
Section~\ref{sec:balls_bins} to obtain a constant-time symmetric algorithm.
\begin{corollary}\label{coro:symmetric_many}
For $l\in \BO(\log n)$, an Algorithm $\A_c(l)$ exists that sends $\BO(l n)$
messages w.h.p.\ and solves Problem~\ref{prob:sbib} with a maximum bin
load of $\BO(1)$ within $\log^*n -\log^* l +\BO(1)$ rounds w.h.p. Balls send and
receive $\BO(l)$ messages in expectation and $\BO(\log n)$ messages w.h.p.
\end{corollary}
\textbf{Proof Sketch.} W.h.p., Algorithm $\A_{\omega}(l)$ assigns coordinators
to a constant fraction of the bins such that these coordinators control
$l_0\in\Omega(l)$ bins. The coordinators inform each of their bins $b$ of the
number of bins $\ell(b)$ they supervise, while any other ball contacts a
uniformly random subset of $l$ bins. Such a bin $b$, if it has a coordinator,
responds with the value $\ell(b)$. Note that the probability that the maximum
value a ball receives is smaller than $l_0$ is smaller than $2^{-\Omega(l)}$;
Theorem~\ref{theorem:chernoff} therefore states that w.h.p.\ 
$(1-2^{-\Omega(l)})n$ balls contact a bin $b$ with $\ell(b)\geq l_0$.

Next, these balls contact the bin $b$ from which they received the largest value
$\ell(b)$. The respective coordinators assign (at most) constantly many of these
balls to each of their bins. By the same reasoning as in
Lemma~\ref{lemma:algo_simple}, we see that (if the constant was sufficiently
large) all but $2^{-\Omega(l)}n$ balls can be placed. Afterwards, we again
proceed as in Algorithm $\A_b^2$, with $k$ initialized to $2^{\alpha l}$ for an
appropriate $\alpha > 0$; analogously to Lemma~\ref{lemma:algo_simple} we obtain
the claimed running bound. The bounds on message complexity can be deduced from
Chernoff bounds as usual.\qed{}

Again, choosing $l=\log^{(r)}n$ for any $r\in \N$, Problem~\ref{prob:sbib} can
be solved within $r+\BO(1)$ rounds using $\BO(n\log^{(r)}n)$ messages w.h.p.

\section*{Acknowledgements}
We would like to thank Thomas Locher and Reto Sp\"ohel.


\pagestyle{empty}

\bibliographystyle{abbrv}
\bibliography{stoc11tech}

%
%
%

\end{document}